\ifdefined\pdfoutput\pdfoutput=1\fi

\documentclass[11pt,table,xcdraw]{article}
\usepackage{EACL2023}

\usepackage[T1,T2A]{fontenc}
\usepackage[utf8]{inputenc}
\usepackage{lmodern}      
\usepackage{times}
\usepackage{latexsym}
\usepackage[T1]{fontenc}
\usepackage{microtype}    
\usepackage{fvextra}
\usepackage{amsmath,amsfonts}
\usepackage{microtype}
\usepackage{inconsolata}

\usepackage{graphicx}
\usepackage{booktabs}
\usepackage{array}
\usepackage{multirow}
\usepackage{subcaption}
\usepackage{caption}
\usepackage{float}
\usepackage{placeins}
\usepackage{colortbl}
\usepackage{stfloats}
\usepackage{adjustbox}

\title{\textsc{ReddiX-NET}: A Novel Dataset and Benchmark for Moderating Online Explicit Services}

\author{
\textsuperscript{*}MSVPJ Sathvik \quad
\textsuperscript{*}Manan Roy Choudhury\textsuperscript{1} \quad
Rishita Agarwal\textsuperscript{2} \quad
Sathwik Narkedimilli\textsuperscript{3} \quad
\textsuperscript{\dag}Vivek Gupta\textsuperscript{1} \\
\textsuperscript{1}Arizona State University \\
\textsuperscript{2}Indian Institute of Technology Guwahati \\
\textsuperscript{3}Indian Institute of Information Technology Dharwad \\
\texttt{\textsuperscript{1}mroycho1@asu.edu}, \texttt{\textsuperscript{\dag}vgupt140@asu.edu}
}

\begin{document}
\maketitle
\def\thefootnote{*}\footnotetext{These authors contributed equally to this work and are recognized as joint first authors with equal contribution.}
\def\thefootnote{\dag}\footnotetext{This author supervised the research and serves as the corresponding author.}
\begin{abstract}
The rise of online platforms has enabled covert illicit activities, including online prostitution, to pose challenges for detection and regulation. In this study, we introduce {\sc ReddiX-NET}, a novel benchmark dataset specifically designed for moderating online sexual services and going beyond traditional NSFW filters, derived from thousands of web-scraped NSFW posts on Reddit, categorizing users into six behavioral classes reflecting different service offerings and user intentions. We evaluate the classification performance of state-of-the-art LLMs (GPT-4, LlaMA 3.3-70B-Instruct, Gemini 1.5 Flash, Mistral 8$\times$7B, Qwen 2.5 Turbo, Claude 3.5 Haiku) using advanced quantitative metrics, finding promising results with models like GPT-4 and Gemini 1.5 Flash. Beyond classification, we conduct sentiment and comment analysis, leveraging LLM and PLM-based approaches and metadata extraction to uncover behavioral and temporal patterns, revealing peak engagement times and distinct user interaction styles across categories. Our findings provide critical insights into AI-driven moderation and enforcement, offering a scalable framework for platforms to combat online prostitution and associated harms.
\end{abstract}

\section{Introduction}

Technology has redefined prostitution, shifting it from traditional settings to online sex work~\cite{hughes2004prostitution}, particularly through paid nude video calls. This evolution presents new challenges in cybercrime, content moderation, and legal enforcement, as regulatory loopholes enable exploitation and illicit transactions~\cite{farley2013online}. Beyond legality, psychological and financial consequences loom large—addiction to such services fuels excessive spending, social isolation, and emotional detachment, distorting real-world relationships. 

The illusion of intimacy fosters unrealistic expectations, while prolonged engagement with explicit content contributes to mental health issues like anxiety, depression, and impulse control disorders. Beyond video calls, online sex work thrives through coded ads, content selling, and exhibitionism, leveraging social media algorithms to amplify engagement and drive compulsive consumption~\cite{abdulla2024mental}~\cite{romans2001mental}. This algorithm-fueled expansion deepens the complexities of digital sex economies, making it a pressing issue for law enforcement, mental health experts, and digital platforms alike~\cite{juditha2021communication}.

Detecting and moderating such content is a significant challenge for online platforms. AI-based moderation tools struggle with classifications, often failing to distinguish between explicit content, suggestive discussions, and legal adult work. Users frequently bypass detection using techniques such as filters, altered camera angles, and coded language~\cite{brown2024policing}~\cite{gahn2024abuse}. Live video content presents an additional challenge, as real-time monitoring is still ineffective~\cite{sunde2022conceptualizing}. Furthermore, the lack of well-labeled datasets limits AI models’ ability to differentiate between various forms of adult content and digital prostitution.

We introduce a novel dataset {\sc ReddiX-NET} designed to detect online sexual services, going beyond traditional NSFW (Not Safe For Work) filters by incorporating transactional cues, coded language, and platform-specific behaviors. Unlike existing datasets, ours leverages multilingual data, evolving online trends, and social media patterns, allowing AI to differentiate between legally explicit content and illicit sex work promotions with greater precision.

LLMs excel at uncovering disguised language and transactional intent that conventional filters miss, making AI-driven moderation far more effective in flagging suspicious activity and preventing financial exploitation. These systems can integrate with social media platforms and digital payment networks, ensuring real-time detection and adapting to emerging trends for continuous content regulation. But these systems also have their own biases and confabulations.

Beyond content moderation, this solution has powerful real-world applications. Cybercrime units can track prostitution networks, financial institutions can detect illicit transactions and fraud, and social media platforms can restrict underage access while enforcing content policies. Mental health organizations can analyze user behaviors to recommend intervention programs, while law enforcement can leverage AI insights to combat illegal activities on digital platforms. With continuous updates and ethical safeguards, this approach ensures a balanced and effective strategy for regulating online sexual services while respecting free speech and privacy rights.

The contributions of the paper are as follows:

    a) As of our knowledge we are the first to develop a dataset on online sexual services.

    b) We present the novel observations and findings from the dataset.

    c) We benchmarked the dataset with SOTA LLMs like Gemini, Mistral, GPT, Llama, and Claude.

\begin{figure*}[t]
    \centering
    \includegraphics[width=1.5\columnwidth]{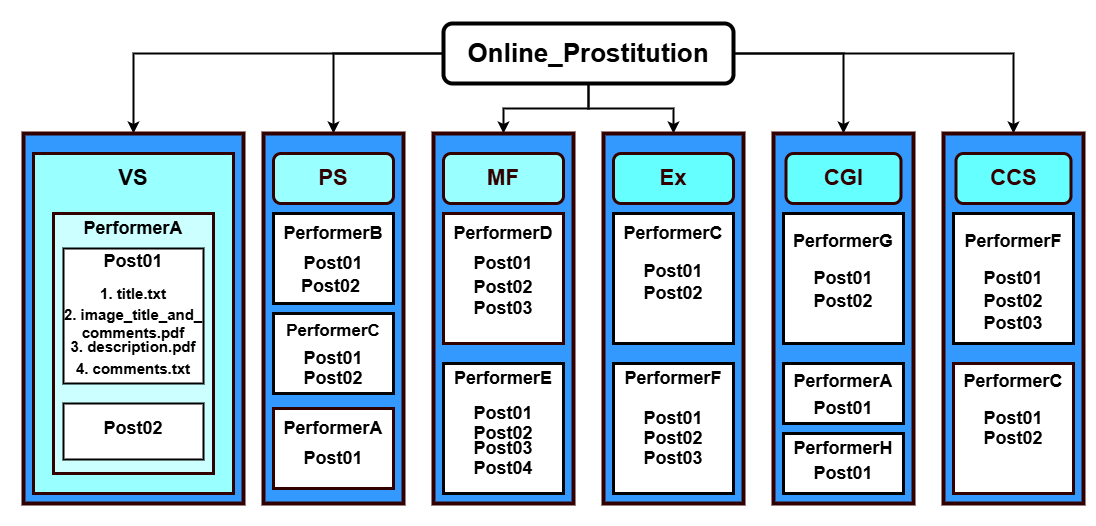}    
    \caption{Structure of the proposed dataset, categorizing online sexual services across six distinct categories and their further subdivisions.}
    \label{taxonomy}
\end{figure*}

\section{Related Work} \label{sec2}

There has been limited research on detecting online sexual services compared to the broader study of online harms. \citet{ibanez2016detecting} introduced a method for identifying US trafficking networks through online escort ads, emphasizing the role of tracking phone numbers for better predictions. Similarly, \citet{keskin2021cracking} analyzed over 10 million ads to identify patterns using text, phone numbers, and images, linking them into circuits and predicting future movements. \citet{ibanez2016detecting} also used content and social network analysis to identify key trafficking indicators and map provider networks. Expanding on this, \citet{giommoni2021identifying} developed software to scrape data from 17,362 UK-based ads, establishing ten human trafficking indicators for quick identification of suspicious cases.

Advanced machine learning and natural language processing have also been applied in this domain. \citet{diaz2020natural} developed a classifier for detecting sex trafficking ads on review sites by training on illegal business data and Yelp reviews. \citet{chopin2023geeks} examined online sex offenders' behaviors, particularly their use of technology for anonymity in child sexual exploitation. Additionally, \citet{wang2020sextraffickingdetectionordinal} proposed an ordinal regression neural network to identify escort ads linked to trafficking, improving lead identification accuracy with a modified cost function and deep learning. Most past work centers on sex trafficking and escort ads, whereas we present a dataset for detecting online sexual services on social media. Our experiments offer key insights into the rise of online prostitution and its impact on today’s generation.

\section{{\sc \textbf{ReddiX-NET}} Construction}\label{sec3}

This section details the methodology and sources used to build {\sc ReddiX-NET}, ensuring a comprehensive and data-driven foundation. It highlights the strategic approach taken to construct a robust and insightful dataset.

\subsection{Data Collection}
The dataset was collected from three large subreddit channels (each with over 50K members) focused on online sexual services with users from all over the world. Due to privacy policies, their names are withheld. Using the Reddit API with PRAW API, we gathered posts offering services like paid meetups, nude video calls, and couple swaps. More details about data collection, cleaning, and preprocessing are provided in Appendix-5. Additionally, to establish the statistical significance of our datasets, we conducted several tests, the results of which are presented in Appendix-6.

\subsection{Data Annotation}
For the experiments, three data annotators, including the authors, carefully reviewed and categorized the services frequently posted by users. The annotation was done meticulously to ensure accuracy in classification. This approach aimed to create a balanced and comprehensive dataset, considering different viewpoints in identifying and categorizing the content effectively.

The posts were divided into the following six categories:

\begin{enumerate}
    \item \textbf{Content Creation and Sales (CCS)}: This category includes services related to the production and sale of adult content, such as videos, photos, or written material, often tailored to specific user requests.
    
    \item \textbf{Couples and Group Interactions (CGI)}: Services in this class involve collaborative engagements, typically between two or more individuals, either for personal interaction.
    
    \item \textbf{Exhibitionism (Ex)}: This category refers to services where individuals perform live or recorded acts for an audience, often emphasizing the act of showcasing themselves in a sexual or provocative context.
    
    \item \textbf{Miscellaneous Fun (MF)}: This class encompasses services that may not fit neatly into the other categories but still involve adult-oriented entertainment or engagement, such as roleplay, fetishes, or casual adult interactions.
    
    \item \textbf{Physical Services (PS)}: Services involving physical interaction, such as in-person meetings, escort services, or any physical contact-based activities are classified here.
    
    \item \textbf{Virtual Services (VS)}: This category includes services provided online, such as video chats, private messages, or virtual performances, which do not involve any physical meetings but are sexual or adult-oriented in nature.
\end{enumerate}


We use defined short forms for categories throughout. Figure~\ref{taxonomy} shows the folder structure with six service-based classes, each containing posts from multiple Reddit performers. Titles, images, and comments reflect engagement, and all classes were manually reviewed for consistency. We measured inter-annotator agreement, achieving high Krippendorff’s alpha ($K$ = 0.69) and Cohen’s kappa ($C$ = 0.66), confirming strong annotation accuracy. Details are in Appendix-5
\subsection{Statistical Analysis}

\FloatBarrier  

A total of 8,146 posts are recorded, with Ex having the highest count (2,302) and MF the lowest (105). The word count across all models reaches 557,764, where Ex leads with 164,131 words, while MF has the least at 10,631. Similarly, comments total 60,240, with Ex having the highest engagement (17,923) and MF the least (1,008). 

\FloatBarrier  
\begin{table*}
\small
    \centering
    \resizebox{\textwidth}{!}{%
    \begin{tabular}{lrrrrrrrr}
    \toprule
    \textbf{Metric} & \textbf{VS} & \textbf{PS} & \textbf{MF} & \textbf{Ex} & \textbf{CGI} & \textbf{CCS} & \textbf{Overall} \\
    \midrule
    No. of Posts per Category                 & 1588   & 1501   & 105    & 2302   & 1103   & 1547   & 8146   \\
    No. of Words per Category                 & 103844 & 102917 & 10631  & 164131 & 73751  & 102490 & 557764 \\
    No. of Comments per Category              & 11373  & 10984  & 1008   & 17923  & 7776   & 11176  & 60240  \\
    Avg No. of Comments per Post per Category & 7.5    & 7.6    & 9.6    & 8      & 7.3    & 7.5    & -      \\
    Avg No. of Tokens per Post per Category   & 72     & 74     & 102    & 75     & 69     & 63     & -      \\
    Avg No. of Posts per Performer per Category & 150  & 162    & 33     & 120    & 97     & 175    & -      \\
    No. of Performers per Category            & 11     & 10     & 3      & 19     & 12     & 9      & -      \\
    \hline
    \end{tabular}}
    \caption{Statistical summary table of the ReddiX-NET dataset, detailing post counts, word counts, comment volumes, and average engagement metrics across the six defined service categories (VS, PS, MF, Ex, CGI, CCS)}
    \label{tab:category-stats}
\end{table*}

Table.~\ref{tab:category-stats} provides an overview of posts, words, and comments across six models: CCS, CGI, Ex, MF, PS, and VS. These statistics highlight the dominance of Ex in content volume and engagement, while MF appears to contribute minimally. This dataset structure provides a foundation for further analysis of content distribution and interaction trends.


         



\section{NLP-Driven Analysis of {\sc \textbf{ReddiX-NET}}} \label{sec4}

This section outlines key experiments on {\sc ReddiX-NET}, using LLMs and PLMs for user classification, sentiment analysis, comment classification, and metadata-temporal analysis. Details follow in subsequent subsections.

\subsection{{\sc \textbf{ReddiX-NET}} User Classification}

This experiment aims to identify users offering specific services based on their posts using LLMs. 
Users employ sophisticated techniques to evade detection, including filters, altered camera angles, and coded language. The dataset is designed to help AI systems recognize these evasion tactics. %
The services provided by users are treated as the ground truth, and our approach uses LLMs to automatically classify these user posts into predefined service categories.

To conduct the classification, we input user posts into various state-of-the-art LLMs and prompt them to categorize the posts into the defined classes. We have some posts which were uncategorized by the LLMs, for those we have used the clustering method to cluster a certain number of posts from the users (based on the needed cluster size) and then tried to classify the whole cluster in one category. For the experimentation, we have utilized several advanced LLM models, including GPT-4\cite{gpt-od}, LLaMA 3.3-70B-Instruct\cite{llama2}, Gemini 1.5 Flash\cite{gemini2023}, Mistral 8$\times$7B\cite{mistral7b}, and Claude Haiku, to assess their performance for this task.

\subsection{{\sc \textbf{ReddiX-NET}} Expression Analysis}
This analysis is done to show how the users respond to the posts and its contents. This helps us understand, how are these posts also affecting the user’s mental health, so that significant measures can be taken based on the psychological impact.

We conducted sentiment analysis on {\sc ReddiX-NET} using both pre-trained and fine-tuned BERT-based models (PLM) and large language models (LLMs) i.e., Qwen 2.5 Turbo, GPT-4o. State-of-the-art LLMs with precision prompts extract nuanced sentiment aspects, including:
(a) Sentiment polarity (positive, neutral, negative, mixed)
(b) Emotional spectrum (joy, anger, sadness, etc.)
(c) Tonal variation (casual, formal, informal, playful, aggressive)
Beyond classification, they provide confidence scores and keyword extraction for deeper insights. A fine-tuned BERT model, trained on domain-specific data, evaluates emotional dependency, state of varied emotions, exploitation, user experience, and mental health concerns. Using Hugging Face’s transformers, BERT maps star ratings to discrete emotions (e.g., "5 stars" to "satisfaction," "1 star" to "aggression"), translating them into societal impact via custom dual-mapping functions.

\subsection{ {\sc \textbf{ReddiX-NET}} Comments Classification}

While sentiment analysis reveals an emotional tone, it doesn’t identify discussion topics in these threads. To address this, we categorized user comments into 19 predefined themes using GPT-4, ensuring accurate classification through contextual understanding.
Unclassified comments undergo a two-step process:
a) Clustering techniques group similar unclassified comments for pattern recognition.
b) LLM-based re-evaluation reassesses and assigns them to the most suitable category then tries to classify them in the existing categories.
This iterative approach enhances classification accuracy, ensuring a structured analysis of user interactions with minimal data loss.

\subsection{Time-based Analysis on {\sc \textbf{ReddiX-NET}}}

We analyzed metadata trends to understand user engagement patterns, focusing on temporal fluctuations in posts and comments. By tracking activity over time, we identified peak and low engagement periods, revealing trends in content consumption and participation. A key focus was on peak hours of user interactions with sexual service-related posts. Analyzing timestamps helped uncover behavioral patterns, user preferences, and content visibility dynamics, offering insights into audience engagement.

\section{Results and Analysis} \label{sec5}

In this section, we will discuss all the results that we have obtained from the experiments that we mentioned in the previous section.
\subsection{{\sc \textbf{ReddiX-NET}} User Classification}

The study evaluates the effectiveness of various large language models (LLMs) in detecting and classifying posts using precision, F1-score, distribution accuracy, precision, errors, and divergence measures (Table.~\ref{tab:performance_extended}).

\begin{table}[ht!]
\centering
\resizebox{0.48\textwidth}{!}{
\begin{tabular}{l c c c c c c}
\toprule
\textbf{Cat.} & \textbf{Pre (\%)} & \textbf{F1 (\%)} & \textbf{MSE} & \textbf{MAE} & \textbf{JSD} & \textbf{Acc (\%)} \\ 
\midrule
\rowcolor[HTML]{FFFFC7} \multicolumn{7}{c}{\textit{GPT-4}} \\
\textbf{VS}   & 45.14 & 61.97 & 0.05 & 0.14 & 0.44 & 85.73 \\
\textbf{PS}   & 39.29 & 56.11 & 0.07 & 0.18 & 0.56 & 81.81 \\
\textbf{MF}   & \textbf{57.14} & \textbf{72.02} & \textbf{0.04} & 0.15 & 0.47 & 84.60 \\
\textbf{Ex}   & 25.00 & 39.58 & 0.06 & \textbf{0.14} & 0.49 & \textbf{85.83} \\
\textbf{CGI}  & 28.57 & 43.15 & 0.13 & 0.20 & \textbf{0.70} & 79.54 \\
\textbf{CCS}  & 33.33 & 49.63 & 0.05 & 0.15 & 0.48 & 84.82 \\ 
\midrule
\rowcolor[HTML]{FFFFC7} \multicolumn{7}{c}{\textit{Llama-3.3-70B-Instruct}} \\
\textbf{VS}   & 49.00 & 64.00 & 0.05 & \textbf{0.14} & 0.51 & 84.71 \\
\textbf{PS}   & 50.00 & 62.20 & 0.05 & 0.15 & 0.51 & \textbf{85.15} \\
\textbf{MF}   & \textbf{68.00} & \textbf{81.00} & 0.08 & 0.20 & 0.60 & 79.83 \\
\textbf{Ex}   & 43.00 & 58.00 & 0.11 & 0.23 & \textbf{0.71} & 77.36 \\
\textbf{CGI}  & 42.00 & 54.00 & 0.11 & 0.20 & 0.67 & 79.85 \\
\textbf{CCS}  & 49.00 & 67.00 & \textbf{0.04} & 0.15 & 0.46 & 85.14 \\ 
\midrule
\rowcolor[HTML]{FFFFC7} \multicolumn{7}{c}{\textit{Mistral 8$\times$7B}} \\
\textbf{VS}   & 42.00 & 56.00 & \textbf{0.03} & \textbf{0.12} & 0.42 & \textbf{87.93} \\
\textbf{PS}   & 47.00 & 61.00 & 0.06 & 0.18 & 0.64 & 82.17 \\
\textbf{MF}   & \textbf{59.00} & \textbf{71.00} & 0.05 & 0.17 & 0.58 & 83.21 \\
\textbf{Ex}   & 40.00 & 55.00 & 0.10 & 0.22 & \textbf{0.70} & 78.43 \\
\textbf{CGI}  & 41.00 & 56.00 & 0.08 & 0.19 & 0.63 & 81.09 \\
\textbf{CCS}  & 44.00 & 58.00 & 0.04 & 0.13 & 0.43 & 86.90 \\
\midrule
\rowcolor[HTML]{FFFFC7} \multicolumn{7}{c}{\textit{Gemini 1.5 Flash}} \\
\textbf{VS}   & 48.00 & 63.21 & 0.05 & \textbf{0.14} & 0.47 & 85.93 \\
\textbf{PS}   & 47.50 & 64.29 & \textbf{0.04} & 0.13 & 0.48 & \textbf{87.24} \\
\textbf{MF}   & \textbf{83.33} & \textbf{83.64} & 0.05 & 0.17 & 0.55 & 82.85 \\
\textbf{Ex}   & 32.50 & 48.21 & 0.06 & 0.17 & 0.61 & 82.70 \\
\textbf{CGI}  & 30.95 & 46.19 & 0.11 & 0.20 & \textbf{0.71} & 80.06 \\
\textbf{CCS}  & 39.52 & 56.08 & 0.05 & 0.16 & 0.48 & 84.34 \\ 
\midrule
\rowcolor[HTML]{FFFFC7} \multicolumn{7}{c}{\textit{Claude 3.5-Haiku}} \\
\textbf{VS}   & 52.00 & 66.00 & 0.06 & 0.16 & 0.54 & 84.08 \\
\textbf{PS}   & 46.25 & 56.85 & 0.05 & \textbf{0.13} & 0.46 & \textbf{86.55} \\
\textbf{MF}   & \textbf{100.00} & \textbf{100.00} & 0.08 & 0.22 & 0.63 & 78.48 \\
\textbf{Ex}   & 35.83 & 51.79 & 0.08 & 0.19 & 0.63 & 81.01 \\
\textbf{CGI}  & 36.67 & 52.38 & 0.13 & 0.21 & \textbf{0.68} & 79.33 \\
\textbf{CCS}  & 37.78 & 52.38 & \textbf{0.04} & 0.15 & 0.46 & 85.49 \\
\bottomrule
\end{tabular}
}
\caption{Comparative evaluation of different large language models (LLMs) across various service categories (VS, PS, MF, Ex, CGI, CCS). The models are assessed based on multiple performance metrics, including Precision (Pre), F1-score, Mean Squared Error (MSE), Mean Absolute Error (MAE), Jensen-Shannon Divergence (JSD), and Accuracy. The category is abbreviated as Cat.}
\label{tab:performance_extended}
\end{table}
The results show varying performance across models, with notable differences in classifying behavioral categories. GPT-4 demonstrated moderate classification performance, excelling in identifying MF (72.02\% F1) and VS (61.97 \% F1) but struggling with Ex (0.39 F1) and CGI (0.43 F1), indicating difficulty in distinguishing nuanced behaviors. While its accuracy ranged between 79\% to 86\%, inconsistencies were evident due to variability in classification confidence. LLaMA-3.3-70B-Instruct improved precision, particularly in MF (81\% F1) and CCS (67\% F1), but faced challenges with explicit and ambiguous content, leading to higher misclassification rates in complex cases despite comparable accuracy to GPT-4. Mistral 8$\times$7B exhibited balanced performance, performing well in MF (71\% F1) but showing inconsistency in CGI (56\% F1) and Ex (55\% F1), with occasional divergence in sentiment-based categories. Gemini 1.5 Flash excelled in MF detection (83.64\% F1) and maintained stable classification across most categories but struggled with CGI (46.19\% F1) and Ex (0.48 F1), revealing difficulties in handling synthetic and explicit content. Claude 3.5 Haiku achieved perfect classification in MF (100\% F1, 100\% precision), outperforming all other models in this category, yet its performance was more moderate in CGI (53\% F1) and CCS (52.38\% F1). Although its accuracy remained high (79\% to 87\%), frequent inconsistencies across multiple categories suggested classification instability.

\subsection{{\sc \textbf{ReddiX-NET}} Expression Analysis}

As discussed in the previous section, understanding the expressions of the user posts is done in two ways i.e., using LLMs and PLM. 
 
\begin{figure}[H]
    \centering
    \includegraphics[width=1\linewidth]{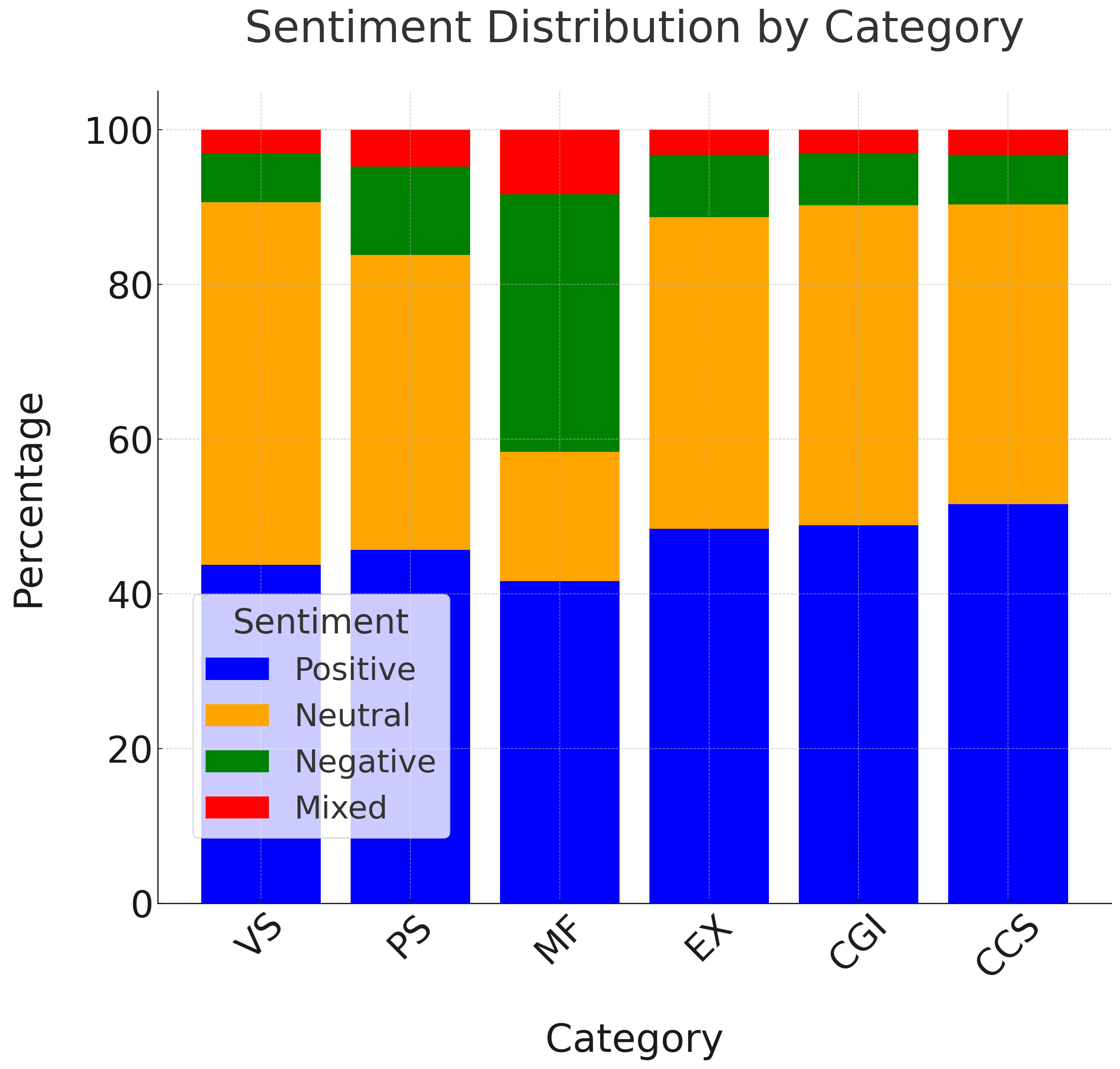}
    \caption{Distribution of sentiment classifications across six different service categories (VS, PS, MF, Ex, CGI, CCS).}
    \label{tab:category-sentiment}
\end{figure}

\textbf{Sentiment Analysis:} First, we will discuss the results of the analysis done using LLMs and then PLMs. The sentiment distribution across service categories given below in Figure.~\ref{tab:category-sentiment}, reveals that EX exhibits the highest overall engagement, with a substantial proportion of negative sentiment, suggesting strong emotional reactions within this category. VS and CCS also show significant activity, with a relatively balanced mix of positive, neutral, and negative sentiments. PS and CGI display moderate engagement, whereas MF has the lowest overall count, indicating minimal interaction in this category.

 The dominance of negative sentiment especially in MF highlights potential concerns or dissatisfaction, while in the other categories, we can notice the dominance of positive sentiments indicating the satisfaction of the users in the other categories' contents.

\textbf{Emotion Type Analysis:} The table below captures the emotions of the users who are engaged in viewing these contents expressed by them in their comments to the post.
\begin{table}[H]  
\centering
\small
\resizebox{\columnwidth}{!}{%
\begin{tabular}{lcccccc}
\toprule
\textbf{Emo(\%)} \hspace{-10pt} & \textbf{VS} & \textbf{PS} & \textbf{MF} & \textbf{EX} & \textbf{CGI} & \textbf{CCS} \\
\midrule
\textbf{Des}  & 31.8  &  25.3  &  8.5  &  32.6  &  27.6  &  31.3  \\
\textbf{Joy}  & 31.8  &  11.4  &  17.1  &  27.2  &  22.7  &  23.7  \\
\textbf{None}  & 7.8  &  0.3  &  7.0  &  11.0  &  15.1  &  10.7  \\
\textbf{Int}  & 9.2  &  1.5  &  3.9  &  9.8  &  10.1  &  9.2  \\
\textbf{Sur}  & 5.6  &  0.1  &  7.8  &  6.0  &  6.5  &  6.9  \\
\textbf{Ant}  & 4.4  &  10.1  &  3.1  &  4.3  &  5.4  &  4.0  \\
\textbf{Lus}  & 2.5  &  16.4  &  4.7  &  3.3  &  2.1  &  4.6  \\
\textbf{Ang}  & 1.8  &  6.3  &  3.1  &  1.4  &  3.2  &  2.9  \\
\textbf{Exc}  & 1.4  &  9.5  &  1.6  &  2.2  &  2.5  &  2.1  \\
\textbf{Ind}  & 1.1  &  0.6  &  7.8  &  0.8  &  1.7  &  0.9  \\
\textbf{Pla}  & 0.8  &  1.4  &  15.5  &  0.1  &  1.0  &  0.8  \\
\textbf{Dis}  & 0.7  &  12.6  &  5.4  &  0.5  &  0.8  &  1.4  \\
\textbf{Inf}  & 0.1  &  0.1  &  12.4  &  0.3  &  1.1  &  1.0  \\
\textbf{Frus}  & 1.0  &  4.4  &  2.3  &  0.5  &  0.2  &  0.5  \\
\bottomrule
\end{tabular}}
\caption{Detailed breakdown percentages of top 14 emotions (Emo) types across six service categories (VS, PS, MF, EX, CGI, CCS). The emotions include Desire (Des), Joy, Interest (Int), Surprise (Sur), Anticipation (Ant), Lust (Lus), Anger (Ang), Excitement (Exc), Indifference (Ind), Playfulness (Pla), Disgust (Dis), Informational (Inf), and Frustration (Frus), along with instances labeled as None.}
\label{tab:category-emotions}
\end{table}

Table.~\ref{tab:category-emotions} Emotional analysis shows high Desire in EX (32.6\%) and VS (31.8\%), with matching Joy levels (27.2\% and 31.8\%), indicating strong emotional engagement. PS stands out with elevated Lust (16.4\%) and Disgust (12.6\%) and minimal Neutral responses (0.3\%), suggesting a provocative or polarizing tone. CGI and CCS show more balanced emotions, with CGI notable for Neutral (15.1\%). MF has the lowest emotional intensity, especially in Joy (17.1\%) and Playfulness (15.5\%). These trends highlight EX and VS as casual and engaging, while PS leans more toward provocative expression.

It is important to note that the above observations are based on a limited and domain-specific dataset; therefore, the interpretations should be viewed as preliminary hypotheses rather than definitive conclusions. Further analysis with larger and more diverse data samples is necessary to validate these findings.
\\
\textbf{Tonality Analysis:} This sub-section examines the linguistic and emotional traits of online sexual service interactions, analyzing tone, emotion, and sentiment distributions across service categories. We identify engagement patterns, explore their correlations, and discuss psychological and social implications, highlighting the interplay between language, affect, and behavior in digital spaces.

\begin{table}[htbp]
\centering
\small
\resizebox{\columnwidth}{!}{%
\begin{tabular}{lcccccc}
\toprule
\textbf{Tone(\%)} & \textbf{VS} & \textbf{PS} & \textbf{MF} & \textbf{EX} & \textbf{CGI} & \textbf{CCS} \\
\midrule
\textbf{Cas}  & 62.3  &  28.6  &  45.3  &  61.8  &  61.0  &  62.8  \\
\textbf{For}  & 13.8  &  8.9  &  6.2  &  14.9  &  16.3  &  13.3  \\
\textbf{Inf}  & 6.9  &  8.9  &  4.7  &  6.4  &  6.3  &  5.5  \\
\textbf{Neu}  & 6.2  &  10.3  &  15.6  &  7.7  &  8.1  &  8.1  \\
\textbf{Play}  & 5.5  &  13.6  &  7.8  &  4.3  &  3.3  &  5.2  \\
\textbf{Agg}  & 1.7  &  4.2  &  6.2  &  1.7  &  2.4  &  2.0  \\
\textbf{Ero}  & 1.4  &  14.5  &  6.2  &  1.3  &  1.2  &  0.9  \\
\textbf{Flir}  & 1.2  &  9.1  &  4.7  &  1.1  &  0.8  &  1.5  \\
\textbf{Se}  & 1.0  &  1.9  &  3.1  &  0.9  &  0.7  &  0.7  \\
\bottomrule
\end{tabular}}
\caption{Illustration of the distribution of top nine tonal expressions across six service categories. The tones include Casual (Cas), Formal (For), Informal (Inf), Neutral (Neu), Playful (Play), Aggressive (Agg), Erotic (Ero), Flirtatious (Flir), and Sexual (Se). }
\label{tab:transposed-tone-distribution}
\end{table}

Table~\ref{tab:transposed-tone-distribution} The percentage-based analysis presents the distribution of tone across service categories. Casual language remains predominant across all categories, with the highest proportions observed in Exhibitionism (EX: 61.8\%) and Virtual Services (VS: 62.3\%). In contrast, Physical Services (PS) exhibits a comparatively higher use of erotic (14.5\%) and playful (13.6\%) tones, emphasizing a strategic use of provocative language. Formal tones, ranging between 13.3\% and 16.3\%, are notably present in VS, EX, Couples \& Group Interaction (CGI), and Content Creation \& Sales (CCS), balancing informal expressions. Aggressive, flirtatious, and sexual/impersonal tones appear infrequently, reflecting the nuanced relationship between service type and linguistic expression.

\textbf{Cross Correlation:}
Figure.~\ref{figure2} (Appendix-2 section) illustrates that the relationship between sentiment and tone is generally weak or inconsistent. A positive sentiment does not invariably correspond to a casual tone, nor does a negative sentiment necessarily imply aggression. In contrast, emotion and tone exhibit stronger correlations; for example, anger aligns with an aggressive tone while joy is more often associated with a playful tone. Moreover, unique patterns emerge across service categories: VS and CGI demonstrate negative sentiment-emotion correlations, PS shows near-zero correlations (suggesting its transactional nature), and MF is the only category where sentiment and emotion are strongly aligned.

\begin{figure}[t]
    \centering
    \includegraphics[width=\columnwidth]{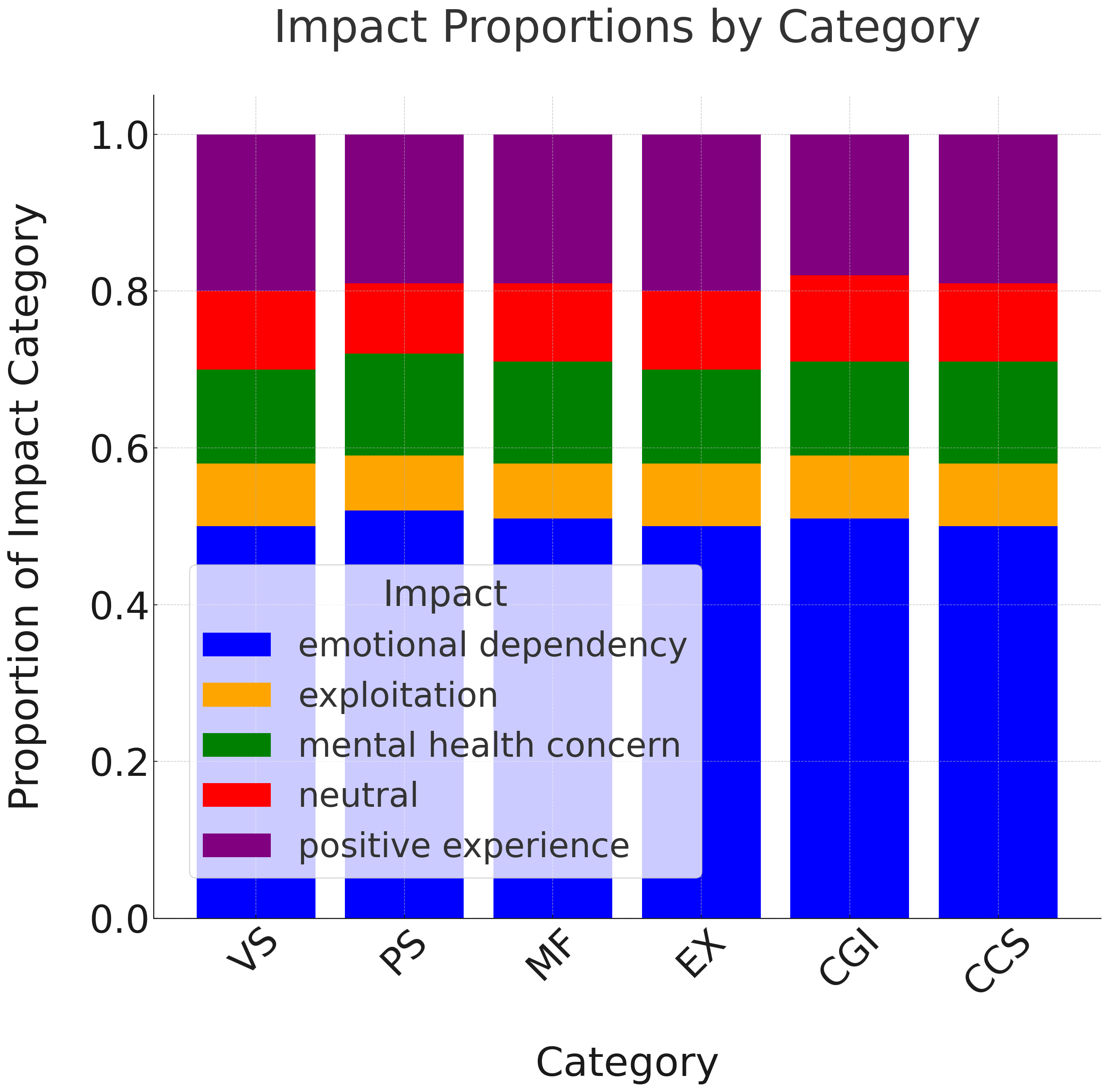}
    \caption{Category-wise impact proportion highlighting the proportion of emotional dependency, exploitation, mental health concerns, neutral perceptions, and positive experiences across different online prostitution categories. This provides insights into the psychological and socio-emotional consequences associated with various engagement types}
    \label{fig:brief_working}
\end{figure}

\paragraph{Psychological and Social Implications:}
Figure.~\ref{fig:brief_working} explores the psychological and social impact on people viewing and engaging in these contents. VS sentiment analysis highlights deep emotional dependency, driving compulsive behaviors, unrealistic expectations, and emotional detachment that strain real-life relationships. It fosters parasocial bonds and addictive consumption, leading to self-esteem issues and body image distortions for consumers, while creators face burnout. Risks of exploitation demand strict regulation to prevent manipulation. Prolonged exposure desensitizes users to real emotional connections, reshaping intimacy and reinforcing unhealthy cycles. Though some report positives, widespread distress calls for urgent intervention strategies. 
While these insights reflect significant patterns observed in our analysis, they remain indicative rather than definitive, shaped by the scope and limitations of our dataset.

\subsection{{\sc \textbf{ReddiX-NET}} Comments Classification}

Comment classification presented in Table~\ref{tab:transposed-table} organizes user interactions based on the nature of their comments or behaviors.

\begin{table}[h!]
\small
\centering
\resizebox{\linewidth}{!}{%
\begin{tabular}{lrrrrrr}
\toprule
\textbf{Buc(\%)} & \hspace{-60pt} 
\textbf{VS} & 
\textbf{PS} & 
\textbf{MF} & 
\textbf{Ex} & 
\textbf{CGI} & 
\textbf{CCS} \\
\midrule
\textbf{\( b_{t_1}\)}  & 0.4  &  0.5  &  0.3  &  0.3  &  0.7  &  0.7  \\
\textbf{\(b_{t_2}\)}  & 0.4  &  0.5  &  1.0  &  0.9  &  1.6  &  1.1  \\
\textbf{\(b_{t_3}\)}  & 0.5  &  0.9  &  1.1  &  1.7  &  1.7  &  1.8  \\
\textbf{\(b_{t_4}\)}  & 0.7  &  1.1  &  1.3  &  0.4  &  1.9  &  1.7  \\
\textbf{\(b_{t_5}\)}  & 1.3  &  1.1  &  2.7  &  4.5  &  3.8  &  3.6  \\
\textbf{\(b_{t_6}\)}  & 1.3  &  1.8  &  1.0  &  1.7  &  0.9  &  0.7  \\
\textbf{\(b_{t_7}\)}  & 0.5  &  0.8  &  1.4  &  0.7  &  1.9  &  0.9  \\
\textbf{\(b_{t_8}\)}  & 1.7  &  2.7  &  2.2  &  2.8  &  4.5  &  6.7  \\
\textbf{\(b_{t_9}\)}  & 12.3  &  11.9  &  9.1  &  6.1  &  3.9  &  4.5  \\
\textbf{\(b_{t_{10}}\)}  & 0.7  &  3.2  &  2.0  &  5.0  &  4.5  &  4.5  \\
\textbf{\(b_{t_{11}}\)}  & 2.7  &  5.5  &  6.0  &  4.5  &  3.2  &  3.1  \\
\textbf{\(b_{t_{12}}\)}  & 5.3  &  4.6  &  14.7  &  3.3  &  6.4  &  3.6  \\
\textbf{\(b_{t_{13}}\)}  & 7.9  &  3.6  &  2.6  &  21.2  &  16.7  &  20.6  \\
\textbf{\(b_{t_{14}}\)}  & 6.2  &  6.4  &  7.7  &  7.8  &  6.4  &  5.4  \\
\textbf{\(b_{t_{15}}\)}  & 1.3  &  2.0  &  5.2  &  2.2  &  9.7  &  5.4  \\
\textbf{\(b_{t_{16}}\)}  & 24.6  &  27.3  &  18.6  &  13.9  &  12.8  &  14.3  \\
\textbf{\(b_{t_{17}}\)}  & 2.6  &  2.3  &  4.0  &  3.4  &  5.2  &  4.5  \\
\textbf{\(b_{t_{18}}\)}  & 19.3  &  19.3  &  13.5  &  10.6  &  7.1  &  8.1  \\
\textbf{\(b_{t_{19}}\)}  & 10.1  &  4.6  &  5.9  &  8.9  &  7.1  &  8.9  \\
\hline 
\end{tabular}}
\caption{This table presents the comments classification of the posts of the different users. Each of the 19 bucket (Buc) types \( b_{t_i} \in \{1,2,\dots,19\} \) captures a distinct user comment or behavior.}
\label{tab:transposed-table}
\end{table}

The baskets/categories are defined as: \( b_{t_{1}}\): Payment or delivery complaints. \( b_{t_{2}}\): Fantasy and violent demands. \( b_{t_{3}}\): Legal and ethical concerns. \( b_{t_{4}}\): Competition or self-promotion. \( b_{t_{5}}\): Emotional support requests. \( b_{t_{6}}\): Unclassified comments. \( b_{t_{7}}\): Price or service negotiations. \( b_{t_{8}}\): Verification and identity inquiries. \( b_{t_{9}}\): Specific content requests. \( b_{t_{10}}\): External link sharing. \( b_{t_{11}}\): Unsolicited requests or harassment. \( b_{t_{12}}\): Reviews and recommendations. \( b_{t_{13}}\): Service demands (intent to purchase/engage). \( b_{t_{14}}\): Skepticism or authenticity questions. \( b_{t_{15}}\): Multilingual comments. \( b_{t_{16}}\): Positive engagement (enjoying the post). \( b_{t_{17}}\): Self-assertive or confident expressions. \( b_{t_{18}}\): Sexual propositions or explicit requests. \( b_{t_{19}}\): Ambiguous or multi-response comments.
 
Exhibitionism (EX) has the highest comment share (21.2\%), followed by Virtual Services (VS, 12.3\%) and Content Creation \& Sales (CCS, 11.9\%). Bucket \( b_{t_{16}} \) (positive engagement) dominates in VS (24.6\%), PS (27.3\%), and EX (13.9\%), indicating strong user interaction. Complaints \( b_{t_{1}} \) and violent demands \( b_{t_{2}} \) are more frequent in EX (0.3\%, 0.9\%), as are legal/ethical concerns \( b_{t_{3}} \) and self-promotion \( b_{t_{4}} \) (1.7\%, 0.4\%). Emotional support \( b_{t_{5}} \) is notable in PS (1.8\%) and EX (4.5\%). Specific content requests \( b_{t_{9}} \) peak in VS (12.3\%) and PS (11.9\%). Service demands \( b_{t_{13}} \) are highest in EX (21.2\%). CGI and CCS show more balanced distributions, while PS and VS excel in positive user interactions (\( b_{t_{16}} \)).

\subsection{{\sc \textbf{ReddiX-NET}} Temporal Analysis}

The data analyzed comes from the real subreddit channels, for privacy policies we have to anonymize the channel names.
The temporal analysis combined with sentiment data indicates not just when users are most active, but when they're exhibiting the most concerning emotional patterns. This can help in identifying at what time frames, it is important to track or check more on the exploitation or evasion of these policies on the social platforms.

\begin{figure}[t]
    \centering
    \includegraphics[width=0.90\linewidth]{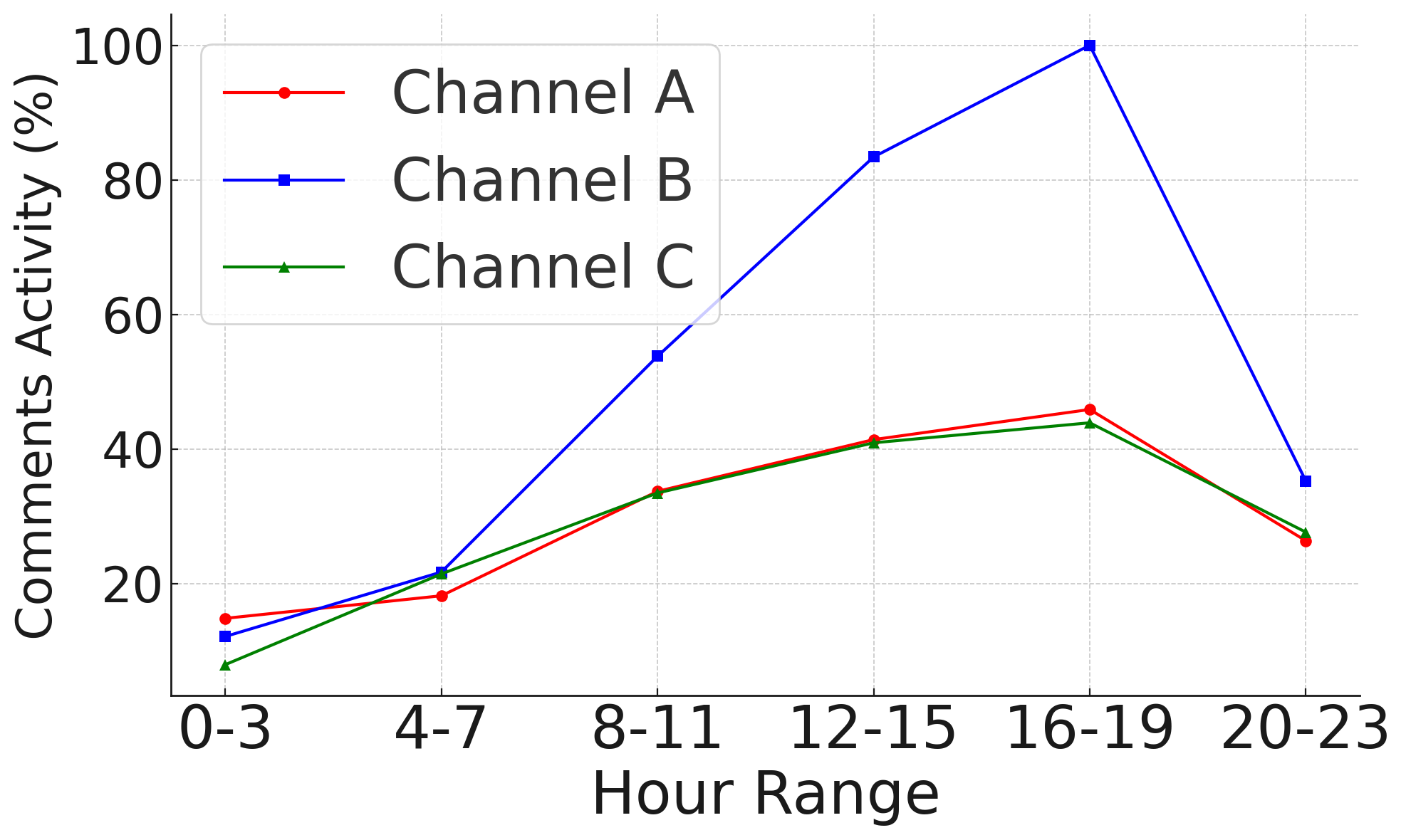}
    \caption{Illustration of comment activity trends over different hourly ranges in a day, highlighting peak engagement times for Channels A, B, and C. The x-axis categorizes the day into six time periods, while the y-axis measures the proportion of total posts within each window.}
    \label{tab:category-sentiment1}
\end{figure}

\begin{figure}[t]
    \centering
    \includegraphics[width=0.90\linewidth]{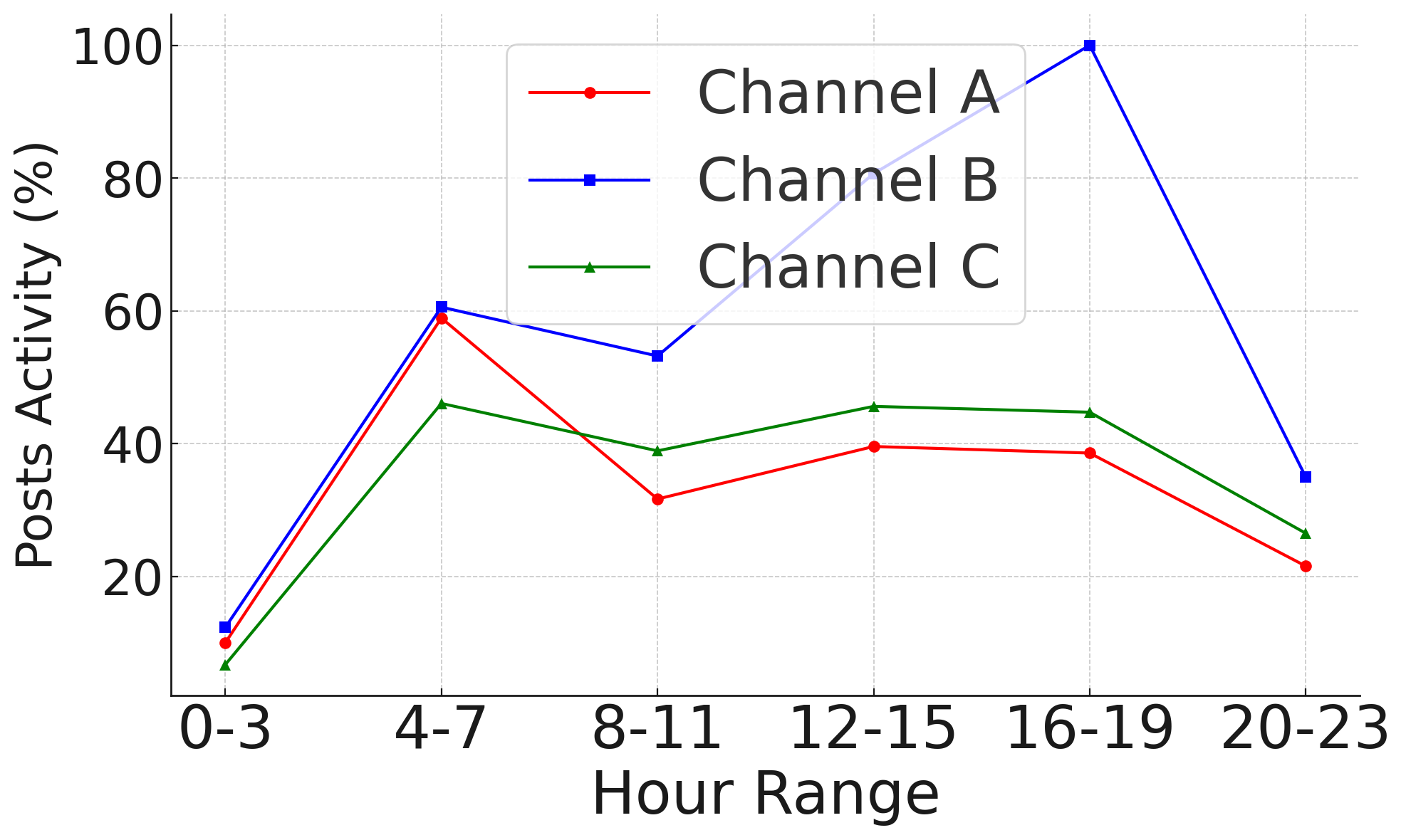}
    \caption{Visualization of temporal posts activity patterns across distinct hourly intervals, emphasizing peak engagement periods for Channels A, B, and C.}
    \label{tab:category-sentiment2}
\end{figure}

\begin{figure*}[ht]
  \centering
  \begin{minipage}[b]{0.32\linewidth}
    \includegraphics[width=\linewidth]{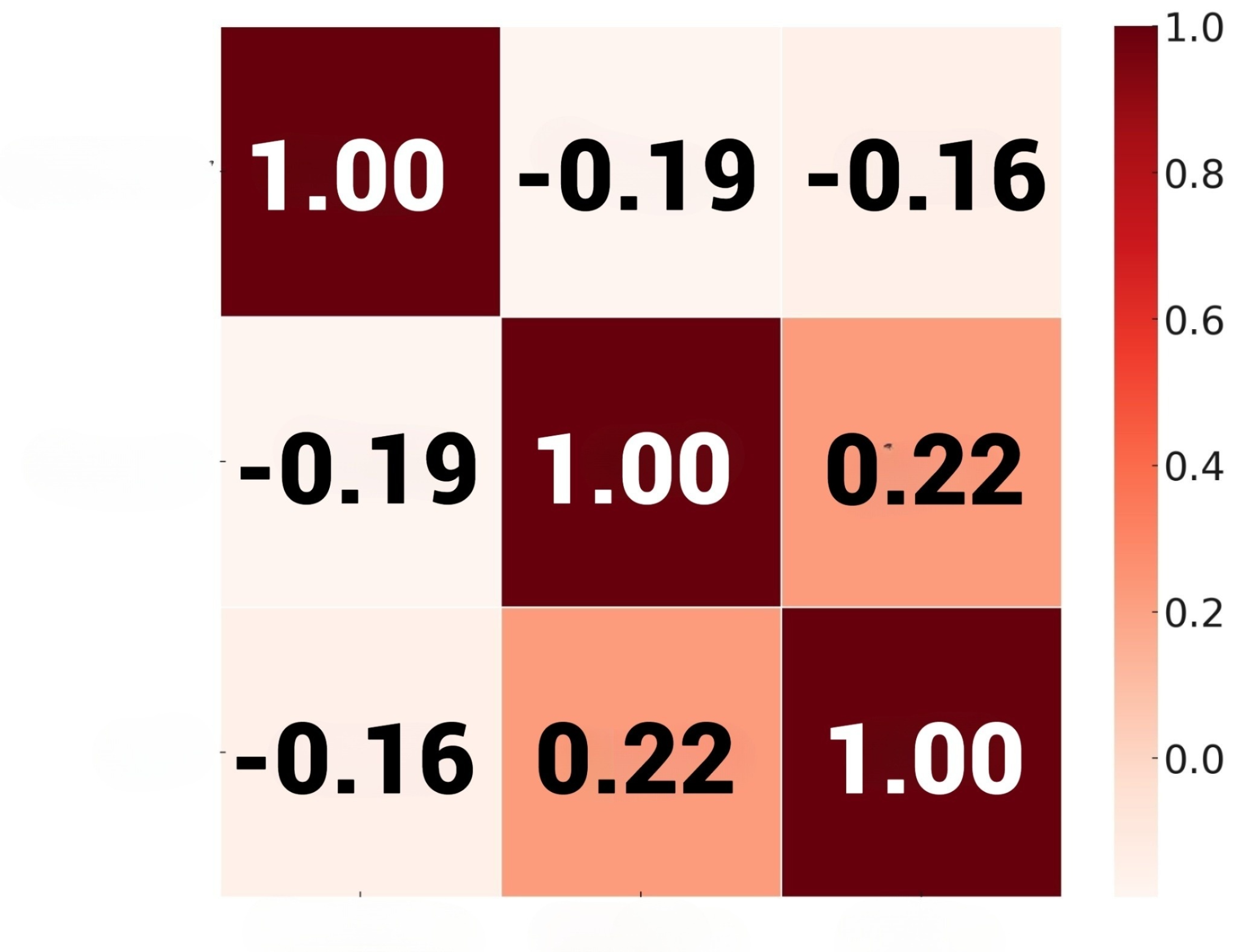}
    \caption*{(a) CCS}
  \end{minipage}
  \hfill
  \begin{minipage}[b]{0.32\linewidth}
    \includegraphics[width=\linewidth]{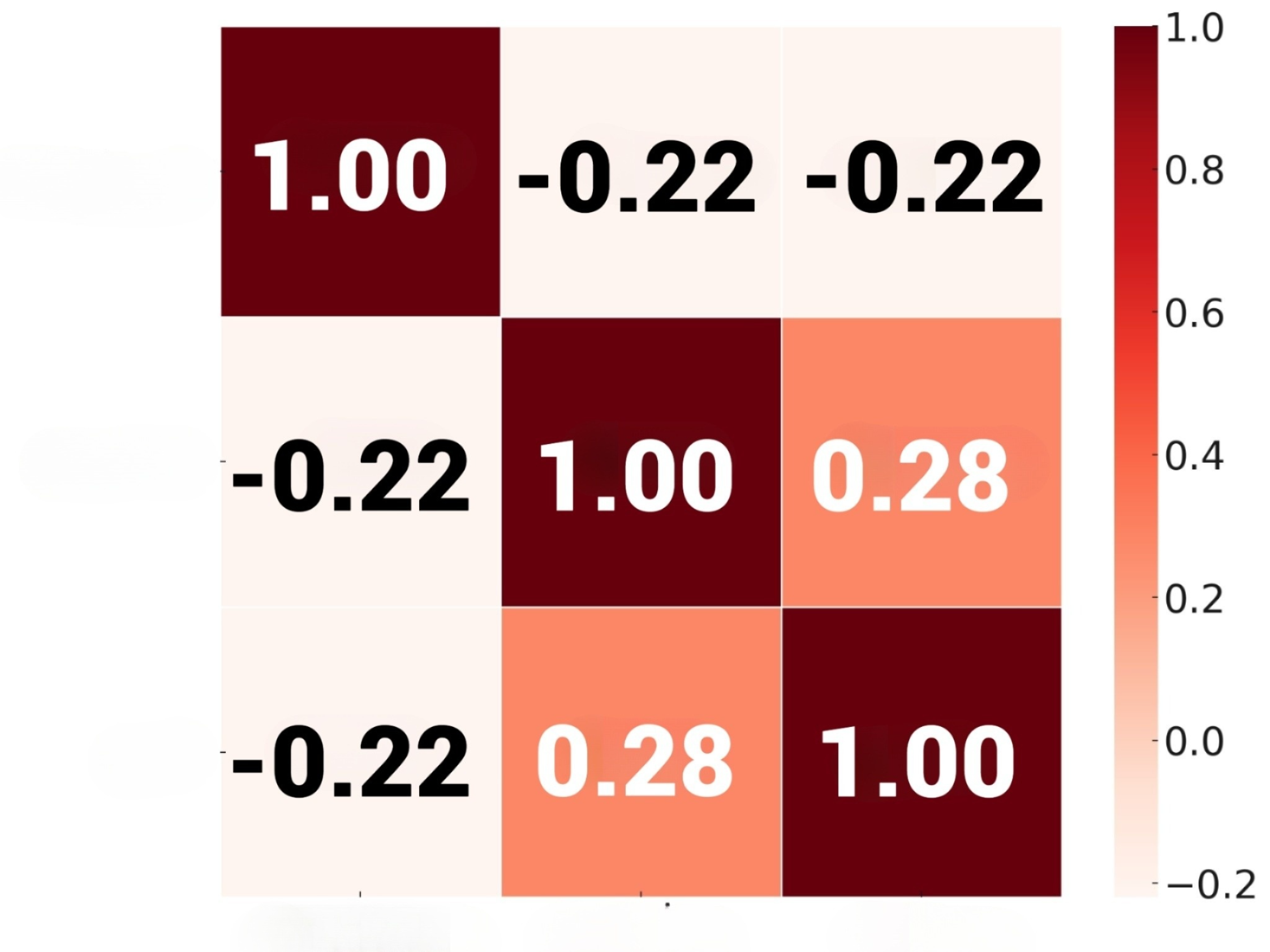}
    \caption*{(b) CGI}
  \end{minipage}
  \hfill
  \begin{minipage}[b]{0.32\linewidth}
    \includegraphics[width=\linewidth]{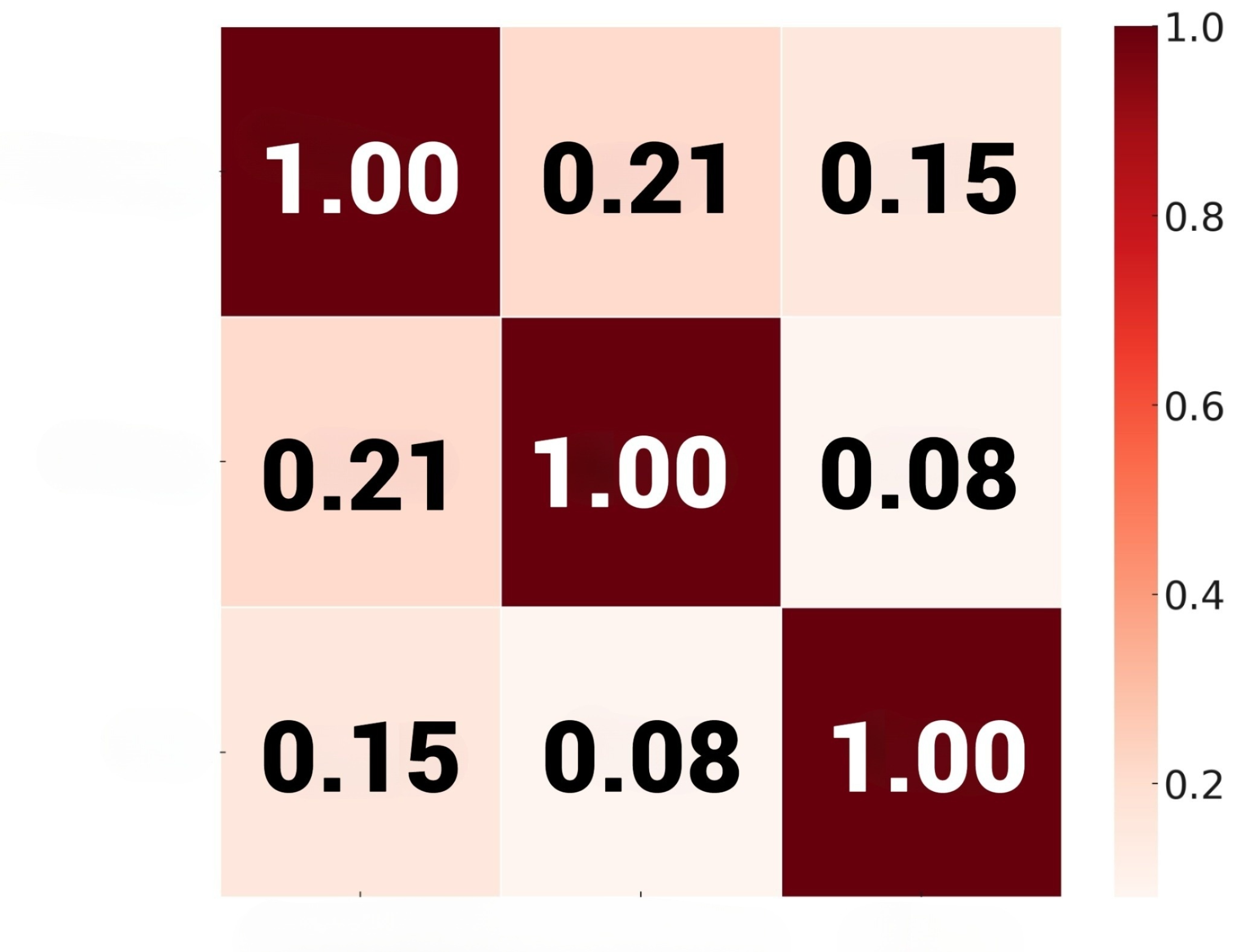}
    \caption*{(c) Ex}
  \end{minipage}
  
  \vspace{0.5cm}
  
  \begin{minipage}[b]{0.32\linewidth}
    \includegraphics[width=\linewidth]{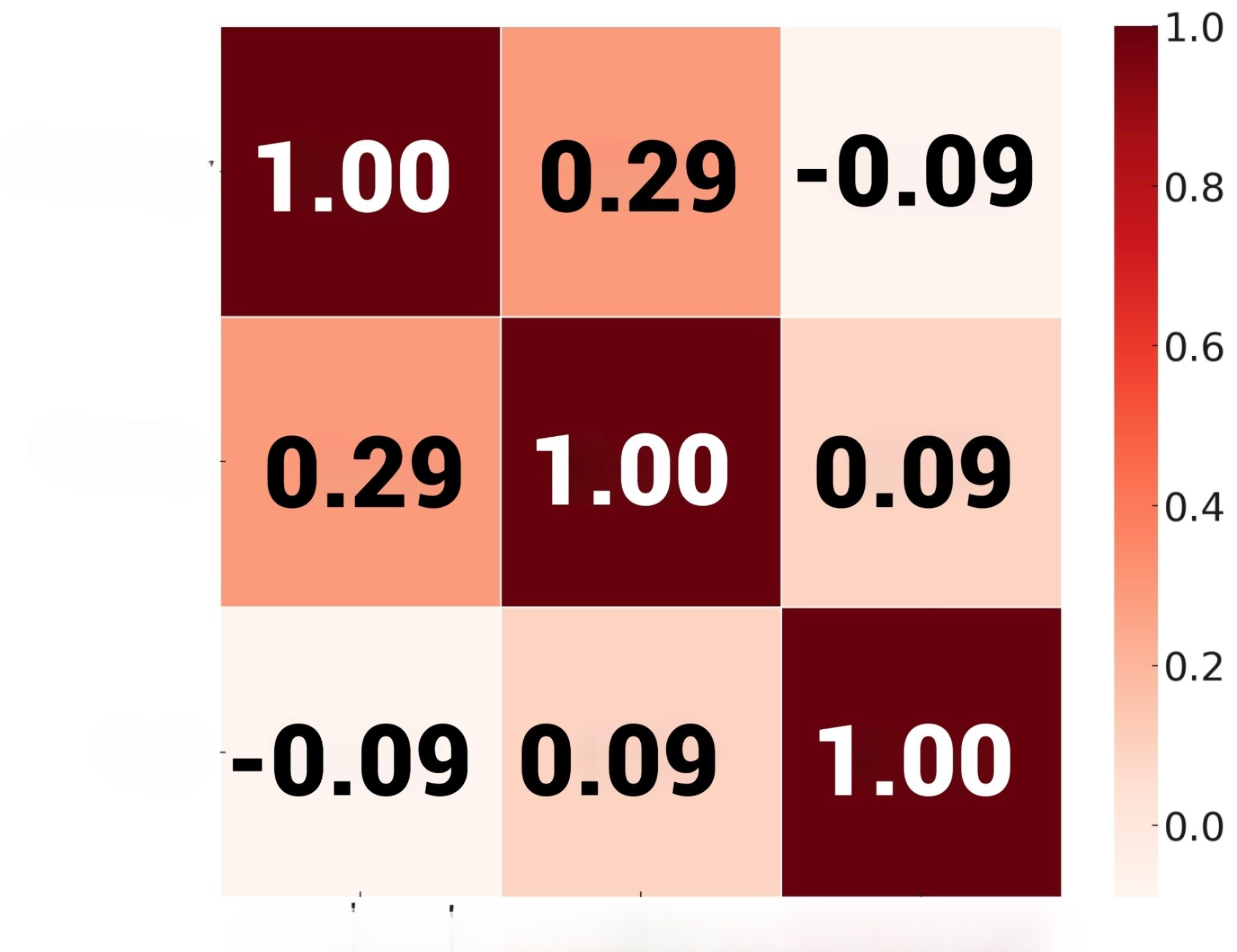}
    \caption*{(d) MF}
  \end{minipage}
  \hfill
  \begin{minipage}[b]{0.32\linewidth}
    \includegraphics[width=\linewidth]{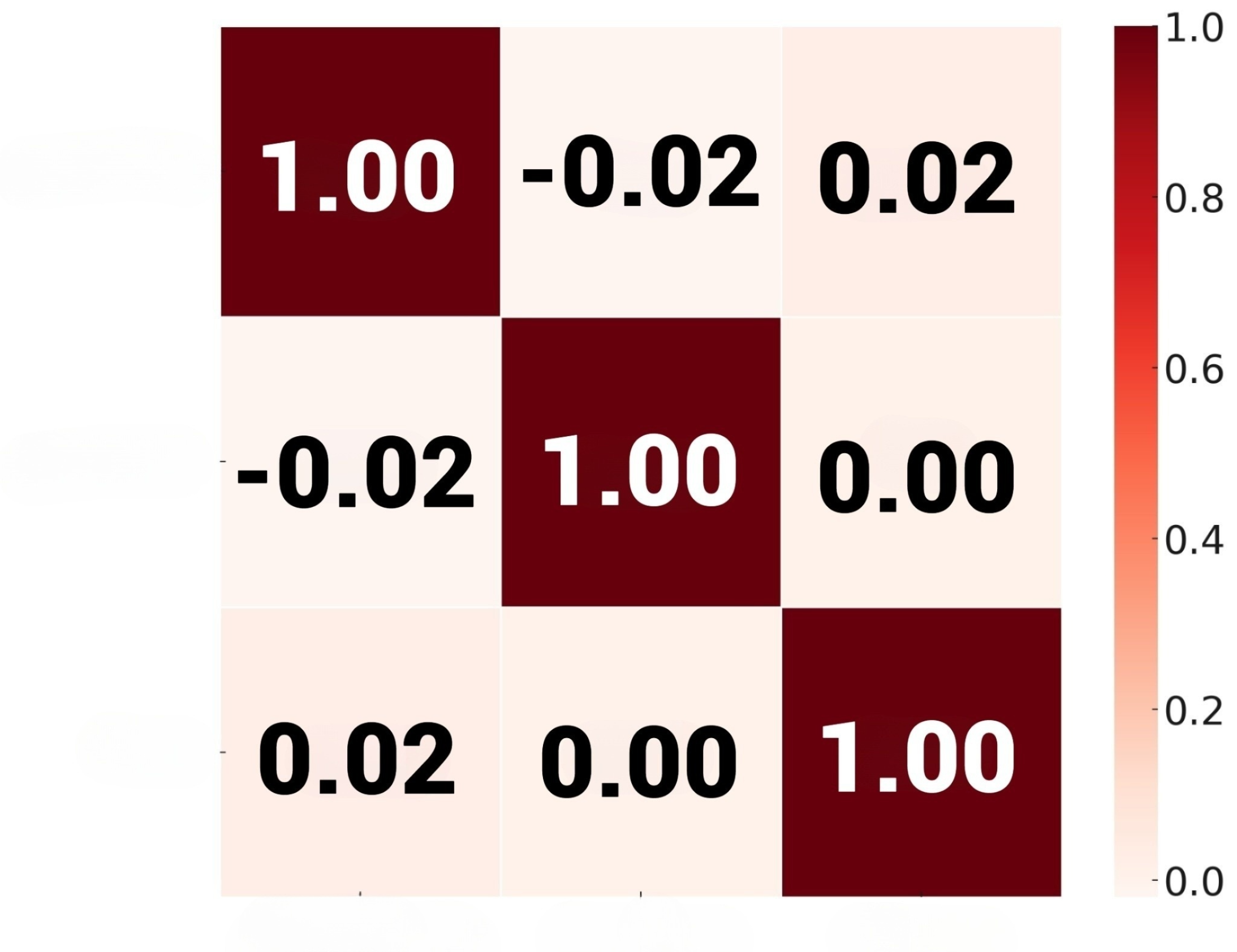}
    \caption*{(e) PS}
  \end{minipage}
  \hfill
  \begin{minipage}[b]{0.32\linewidth}
    \includegraphics[width=\linewidth]{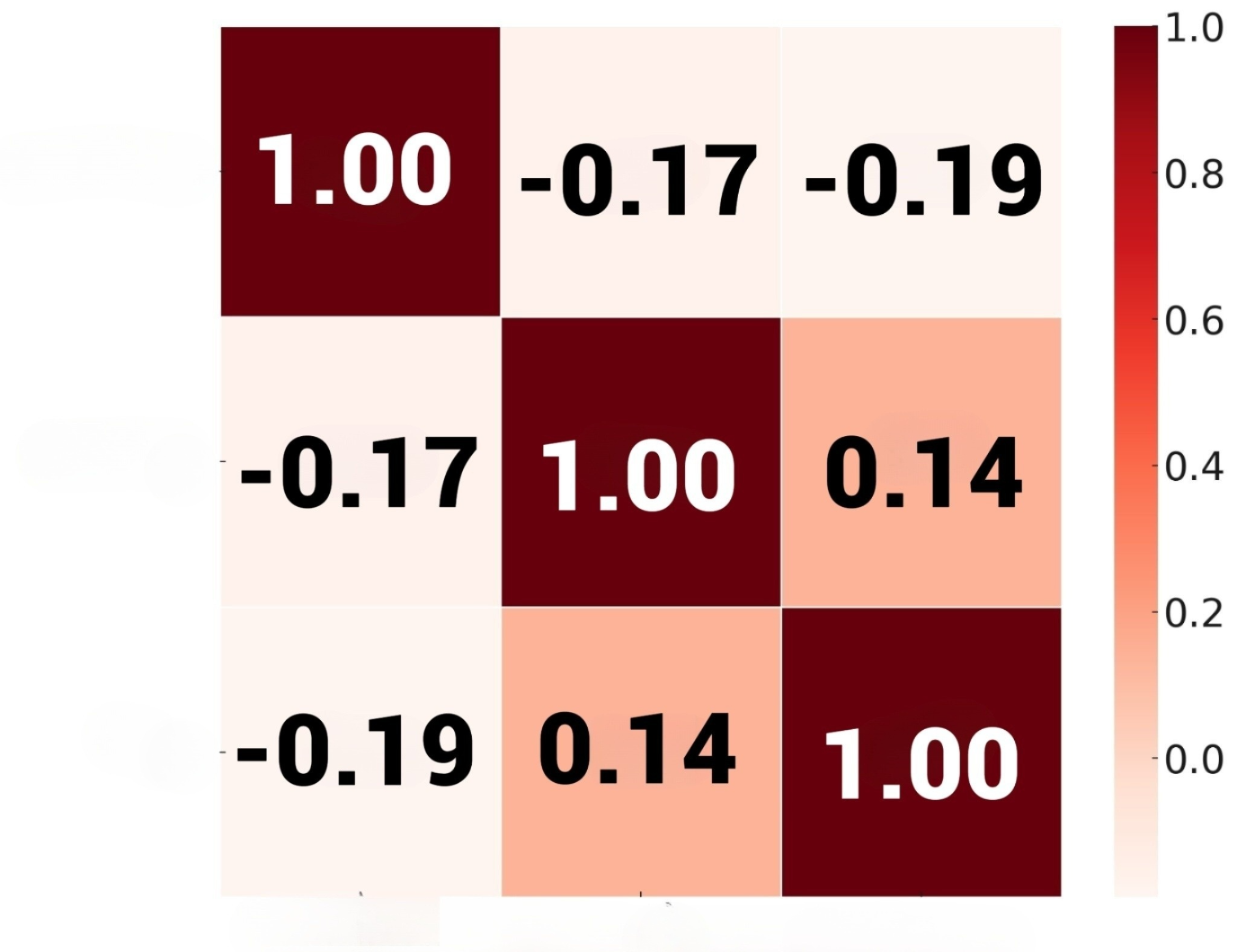}
    \caption*{(f) VS}
  \end{minipage}
  
  \caption{Correlation heatmaps depicting the interplay between sentiment, emotion, and tone across multiple dataset categories. Darker shades denote stronger correlations, while lighter hues indicate weaker association 
  }
  \label{figure2}
\end{figure*}

The graphs illustrate variations in activity levels throughout the day, highlighting specific time periods when user interactions are more frequent. Figure.~\ref{tab:category-sentiment1} and Figure.~\ref{tab:category-sentiment2} reveal a distinct daily cycle in online prostitution discussions on Reddit, with activity peaking between 12-19 hours as both posts and comments surge. 


Engagement starts moderate, rises sharply from 4–11 hours, and peaks in the afternoon/evening, reflecting higher activity during private, non-working hours. Activity drops after 20 hours. A high comment-to-post ratio suggests active participation and aligns with anonymous, leisure browsing. Channel B shows the highest and most consistent engagement, highlighting behavioral and privacy-driven interaction patterns.

We also conducted a detailed class-by-class error analysis, presented in Appendix-4. Additionally, we explored ensemble methods by aggregating different LLMs and observed particularly intriguing results with triadic combinations, as outlined in Appendix-7. Furthermore, we performed an ablation study on various feature importance—focusing on sentiment, emotion, and tone—which yielded valuable insights detailed in Appendix-8.

\section{Conclusion}

This study presents {\sc ReddiX-NET}, a benchmark dataset for detecting and moderating online prostitution services using Reddit data. By categorizing user interactions into six behavioral classes and leveraging state-of-the-art LLMs, we offer a scalable framework for AI-driven moderation. Our findings highlight key challenges in automated detection, including evasion tactics, contextual complexity, and AI moderation limitations. Sentiment and comment analysis further reveal distinct engagement behaviors, underscoring psychological, social, and regulatory implications.
Future research directions include enhancing sentiment analysis through more deterministic methods, integrating AI systems with human-in-the-loop approaches for improved contextual understanding, and investigating the potential mental health impacts of emotionally charged content. Our study also offers several important real-life insights, which are discussed in detail in Appendix-9.

\section*{Limitations}
This study, while offering a novel dataset and benchmark for online sexual service moderation, is subject to certain limitations. Contextual ambiguity in online discussions makes classification difficult, even for advanced LLMs. While we employed diverse annotation strategies, human bias may still affect labeling. The dataset is Reddit-specific, limiting generalizability to other platforms. Additionally, evolving evasion tactics pose ongoing challenges for AI moderation, requiring frequent updates.
\section*{Ethics Statement}

The ethical dimensions of research concerning online sexual services are multifaceted and deeply sensitive, demanding rigorous safeguards and deliberate ethical oversight. Our study seeks to enhance online safety and inform content moderation strategies while remaining acutely aware of the potential for misuse, exploitation, or harm. To uphold the highest ethical standards, we established a comprehensive framework emphasizing participant privacy, robust data security, and principled usage throughout the research lifecycle.

To promote transparency and responsible engagement, we will release a detailed datasheet alongside the \textsc{ReddiX-NET} dataset. This datasheet outlines the dataset’s structure, data collection pipeline, annotation methodology, and usage limitations. We explicitly state that \textsc{ReddiX-NET} is intended strictly for benchmarking and not for model training or any application that could facilitate harm or exploitation.

We prioritized annotator safety and well-being by collaborating with a technical, community-driven organization to recruit and manage our annotation team (One of the authors is a part of this organization). This organization (due to anonymity considerations, we are unable to disclose the name of the organization. However, upon request, all the relevant details will be shared with authorized personnel.) independently oversaw ethical review procedures and secured formal IRB approval for the study. One female annotator was intentionally included in the process to ensure sensitivity toward gender dynamics and to mitigate implicit biases in the annotation of content related to online sexual services. All annotators were clearly informed of the emotionally sensitive nature of the data and were provided with mental health resources and protocols for emotional self-care. Annotators were encouraged to take breaks and access professional support as needed throughout their work.

We implemented a layered and automated anonymization protocol to ensure the complete de-identification of users and channels. All usernames, profile links, subreddit identifiers, timestamps, and geolocation metadata were stripped or replaced with generic placeholders such as \texttt{[USER]}, \texttt{[PROFILE]}, or \texttt{[SUBREDDIT]}. Personally identifiable information (PII), including phone numbers, email addresses, and physical locations, was redacted or tokenized as \texttt{[INFO REDACTED]}. Images were processed through automated pipelines to blur or crop identifiable regions, removing any visual cues to re-identification. A secondary, manual verification step will precede any public release to ensure comprehensive anonymization compliance.

The dataset was sourced from Reddit, a publicly accessible platform, and all data collection adhered to Reddit’s terms of service. We emphasize that despite these precautions, both Reddit and the LLMs used may carry inherent biases. Future users are encouraged to critically evaluate and mitigate such biases in their downstream applications.

\textsc{ReddiX-NET} will be distributed under a restrictive Creative Commons Attribution-NonCommercial-NoDerivatives 4.0 (CC BY-NC-ND 4.0) license, requiring users to accept an ethical usage agreement. Access will be granted only after submission and approval of a valid research use case through a dedicated form. We release only a curated subset of the dataset, not the full corpus, to minimize risks. Additionally, all dataset access will be logged, ensuring accountability and traceability in case of any ethical breaches or re-identification attempts.

To further reinforce ethical use, we outline a responsible usage checklist:

\begin{itemize}
    \item \textbf{Permitted Uses:} Academic, non-commercial research with recognized IRB or ethical committee approval; development of harm-reduction strategies; bias detection and fairness audits; and responsible studies involving sensitive content.
    \item \textbf{Prohibited Uses:} Attempts to re-identify users or profiles; any activity contributing to sexual exploitation or violating relevant legal frameworks such as the Immoral Traffic (Prevention) Act (1956); or commercial use without explicit written approval.
\end{itemize}

A clarification is warranted regarding the annotation team: the annotation effort was directly coordinated by one of the authors, a member of the IRB granting organization's research team. All annotators were also members of this organization, and the work was conducted collaboratively with the research team. Although no financial compensation was provided, the annotators participated voluntarily as they were driven by their strong belief and commitment in the project's objective. The annotators were fully informed of the study’s goals and aligned with its ethical standards under the approved IRB protocol.

By adhering to these principles, we aim to enable responsible research that promotes societal benefit, respects individual dignity, and avoids infringement upon consensual adult expression. This project exemplifies our commitment to ethical innovation and accountable AI deployment in sensitive domains. We acknowledge the use of AI assistants for drafting portions of the paper and supporting related tasks such as editing and formatting, with all content reviewed and finalized by the authors.


\clearpage
\onecolumn
\section*{Appendix-1: Prompts}

\subsubsection*{Hyperparameters}

Key hyperparameters That we have used in our experiments include the temperature, which controls the randomness of predictions, typically set between 0.2 and 0.5 to ensure a more deterministic setting, and the max token length, chosen based on the average post length. Additionally, the models are fine-tuned using a learning rate range of 1e-5 to 1e-3 and a batch size between 16 and 64, with the number of training epochs determined by the convergence of the loss function. During inference, models may use nucleus sampling (top-p) with a probability threshold of 0.9. The evaluation metrics for this task include Distribution Accuracy, Accuracy, F1 Score,  Precision, Mean Absolute Error (MAE), and Shannon Entropy of the distribution.

\subsubsection*{Classification/Evaluation Prompt}

\label{sec:prostitution-categorization}
\small
\begin{Verbatim}[breaklines=true]
Here is the content of a post. It has the following attributes:
1. Title: "{title}"
2. Image Description: "{image_description}"
3. Comments: "{comments}"

Your task is to classify this post into one of the following six categories of services related to online prostitution:
1. Physical Services: Posts offering in-person sexual 
services or physical prostitution.
2. Virtual Services: Posts offering virtual 
interactions such as video calls, virtual sex, or promoting platforms like OnlyFans.
3. Exhibitionism: Posts showcasing exhibitionistic 
behavior, such as public displays or other forms of showcasing oneself.
4. Content Creation and Sales: Posts promoting or 
selling photos, videos, or other content without direct interaction.
5. Couples and Group Interactions: Posts seeking 
interactions involving couples, threesomes, or group scenarios.
6. Miscellaneous Fun/Exploration: Posts describing non-specific fun, exploration, or interactions that do not fall into the other categories.

Carefully analyze the title, image description, and comments. Then, determine which of the six categories the post best fits into. Respond with only the category name. If the information is insufficient, respond with 'Uncategorizable'.
Return only the category name or "Uncategorizable" based on your analysis, in the response.
\end{Verbatim}


Prompt. 1 instructs an AI to classify a post into one of six distinct categories related to online prostitution services based on the post's title, image description, and comments. It delineates clear definitions for each category—ranging from physical and virtual services to exhibitionism, content creation, couples and group interactions, and miscellaneous fun/exploration—ensuring that the classification process is both structured and comprehensive. The prompting strategy emphasizes a careful, contextual analysis of the provided attributes and mandates that the AI return only the appropriate category name or "Uncategorizable" if the information is insufficient, thereby promoting precise and deterministic decision-making in the classification process.

\subsubsection*{Sentiment Analysis Prompt}

\label{sec:sentiment-analysis}
\small
\begin{Verbatim}[breaklines=true]
You are an expert AI performing sentiment analysis.

Analyze the following text and provide the following insights:
    1. Sentiment: Positive, Neutral, or Negative, with a confidence score (0-1).
    2. Emotion Classification: Identify the dominant emotion (e.g., joy, anger, sadness, surprise, etc.).
    3. Keywords: Extract the main keywords or phrases relevant to the context.
    4. Tone: Determine the tone (e.g., formal, casual, playful, persuasive, etc.).

\end{Verbatim}

\textbf{Output Format:}

\begin{itemize}
    \item Sentiment: [label], Confidence: [score]
    \item Emotion: [emotion]
    \item Keywords: [keywords]
    \item Tone: [tone]
\end{itemize}

The provided sentiment analysis prompt instructs an expert AI to perform a comprehensive evaluation of a given text by extracting multidimensional insights. Specifically, it requires the AI to determine the overall sentiment—positive, neutral, or negative—while providing a confidence score, classify the dominant emotion (such as joy, anger, or sadness), extract key phrases or keywords pertinent to the context, and assess the tone (e.g., formal, casual, or playful) of the text. The prompt also specifies a structured output format, ensuring that the results are returned in a consistent and standardized manner. This prompting strategy is designed to facilitate a detailed, context-aware analysis that leverages both qualitative and quantitative dimensions, thereby enhancing the interpretability and reliability of the sentiment analysis process.

\subsubsection*{Comments Classification Prompt}

\label{sec:content-analysis}
\small
\begin{Verbatim}[breaklines=true]
You are an expert content analyst. For each comment provided, classify it into exactly one of the following 18 classes:

1. Users Who Are Enjoying the Post and Its Contents (Engagement & Positive Sentiment)
   - "Wow! You look absolutely stunning!"
   - "Absolutely mesmerizing, I'm hooked!"
   - "Stunning visuals, keep up the great work!"

2. Users Who Are Demanding Such Services (Intent to Purchase/Engage)
   - "How much do you charge for this service?"
   - "Where can I reach you for more details?"
   - "Are you available for a private session?"

3. Users Who Are Requesting Specific Content (Content Demand Trends)
   - "Can you do a video in a red dress?"
   - "I’d love to see more dance moves from you!"
   - "Could you post more outdoor shoots?"

4. Users Who Are Skeptical or Questioning Authenticity (Trust & Credibility Issues)
   - "Is this actually you or just edited?"
   - "Has anyone actually met her? Looks fake."
   - "Seems too polished—are these authentic?"

5. Users Who Are Providing Reviews & Recommendations (Word-of-Mouth & Service Feedback)
   - "She’s super professional and amazing to work with!"
   - "Had a great time, she’s very professional!"
   - "Overpriced, not worth it."

6. Users Who Are Discussing Legality & Ethics
   - "Isn’t this kind of thing banned here?"
   - "Should this even be allowed on this platform?"
   - "I’m concerned about the legality of this content."

7. Users Who Are Competing or Self-Promoting Services
   - "Check out my profile if you like this!"
   - "I offer exclusive content at a discount!"
   - "I can do this for half the price. DM me!"

8. Users Who Are Negotiating Prices or Services
   - "Can you do this for \$40 instead?"
   - "Is there any discount if I book multiple sessions?"
   - "How about a special rate for returning customers?"

9. Users Who Are Complaining About Payment or Delivery Issues
   - "I paid but never got my order!"
   - "She stopped replying after I sent the payment!"
   - "This is a scam, don’t fall for it!"

10. Users Who Are Making Unsolicited Requests or Harassment
    - "Send me something for free first!"
    - "I’ll find you if you don’t reply!"
    - "Do this for me, or else!"

11. Comments that are Multi-Lingual
    - "Teri nazaron mein vo jadu hai."
    - "Include comments from languages which are not English."
    - "acha hai"

12. Comments that are Fantasy and Violent Demands
    - "Show me an online act that blends erotic fantasy with a violent edge."
    - "I demand you to enact a dark fantasy scene with intense aggression."

13. Comments that are Emotional Support
    - "Your posts always brighten my day!"
    - "I appreciate your openness; it helps me feel less alone."

14. Comments that are Verification and Identity Inquiries
    - "Is this really you or just an impersonator?"
    - "Can you prove that this is your real account?"

15. Comments that are External Link Sharing
    - "Check this link out for exclusive content: [external link]"
    - "Visit my page for more: [link]"

16. Comments that are Sexual Propositions / Explicit Requests
    - "Can you send me private pics? I’ll pay extra."
    - "Do you do custom videos with nudity?"

17. Comments that are Self-Assertive/Confidence Expressions
    - "I’m the best at what I do, no one compares!"
    - "I always get what I want, and this is no different."

18. Comments that are Ambiguous or Multi-Response Comments
    - "I’m not sure what to think about this..."
    - "Interesting... I wonder what’s really going on."

For each comment provided, classify it into exactly one of the above categories and return the output as a JSON object with each original comment as a key and its classification as the value.
\end{Verbatim}
The above prompt positions the AI as an expert content analyst, utilizing a persona-based approach to accurately classify user comments into one of 18 predefined categories. This structured prompt incorporates few-shot examples for each category—ranging from positive engagement and service inquiries to external link sharing and ambiguous expressions—thereby guiding the AI with concrete instances of desired outputs. By instructing the AI to evaluate each comment and return a JSON object mapping the original comment to its respective classification, the strategy leverages contextual cues and demonstration-based learning to ensure consistency and precision in categorization.

\clearpage
\section*{Appendix-2: Sentimental Analysis}

\subsection*{Sentimental Analysis using LLMs}

\begin{figure*}[ht]
  \centering
  \begin{minipage}{0.8\linewidth}
    \centering
    \includegraphics[width=\linewidth]{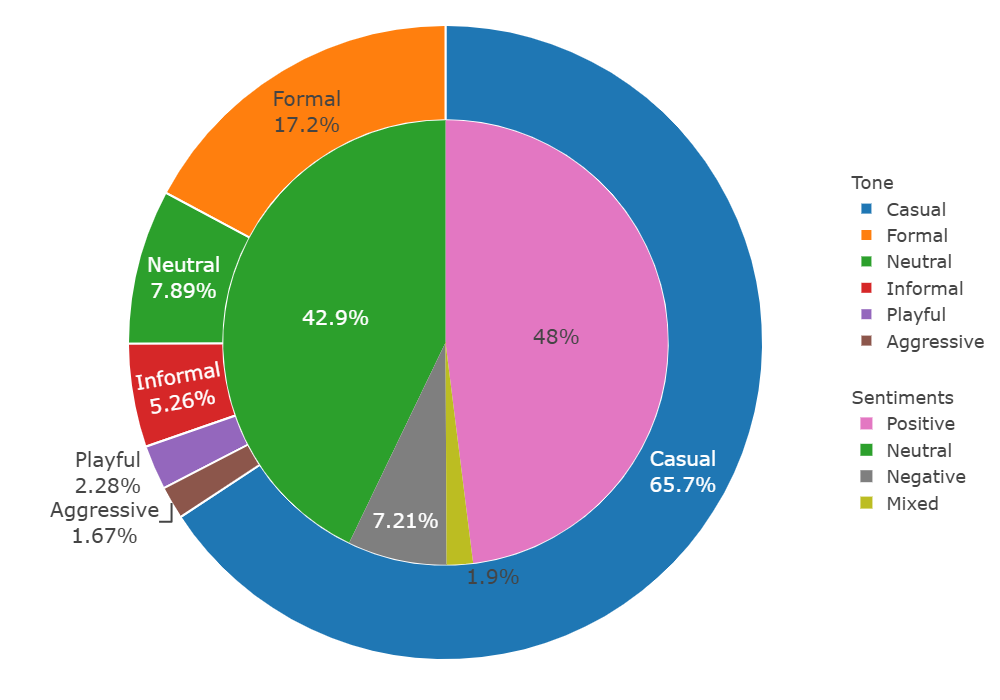}
    \caption*{(a) Sentiment and Tones Distribution across sentiment types for different classes in the dataset for CGI Category.}
  \end{minipage}
  
  \vspace{0.5cm}
  
  \begin{minipage}{0.8\linewidth}
    \centering
    \includegraphics[width=\linewidth]{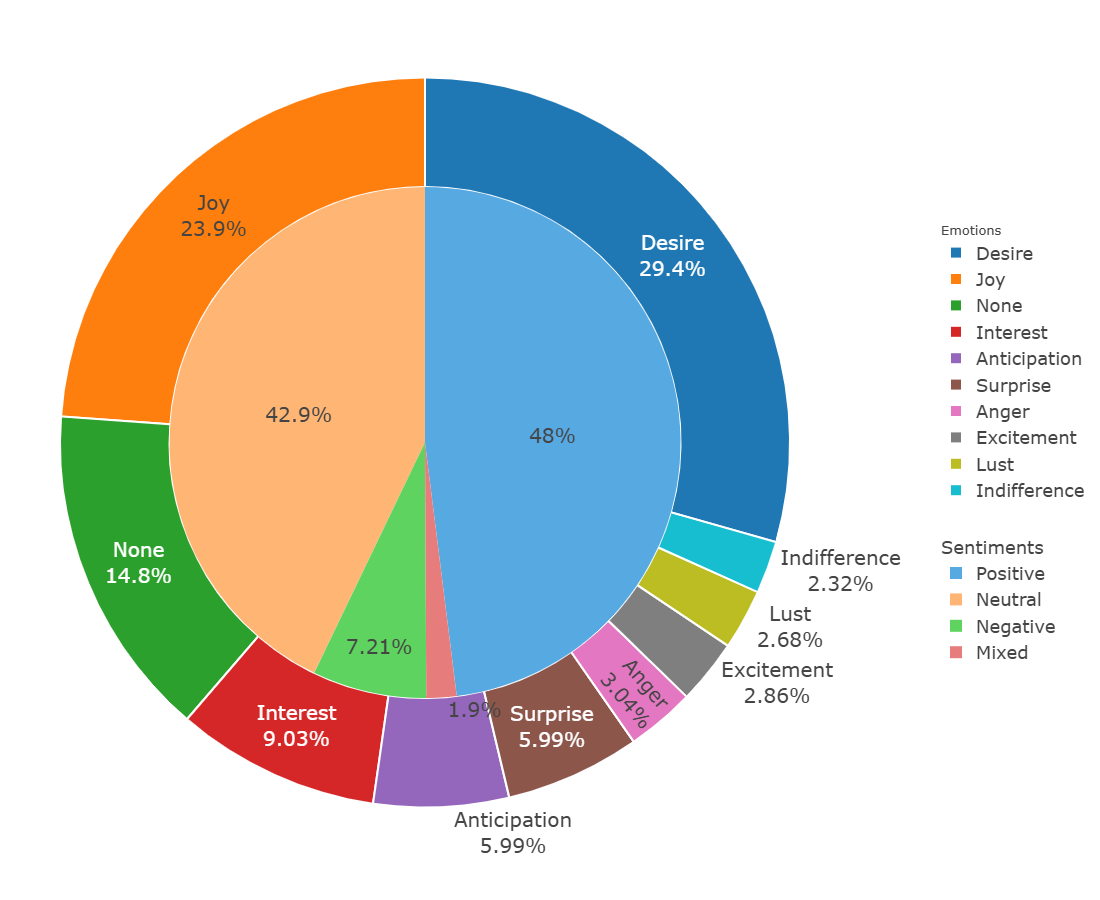}
    \caption*{(b) Sentiment and Emotions Distribution across sentiment types for different classes in the dataset for CGI Category.}
  \end{minipage}
  
  \caption{Distribution of tone and emotions across sentiment types for different classes in the dataset.}
  \label{figure3}
\end{figure*}

\textbf{Distribution of Tone Types Across Sentiment Types:}
Based on the two nested pie charts showing sentiment analysis alongside emotions and tone distribution, here's a comprehensive analysis:
The sentiment distribution in both charts reveals a predominantly positive and neutral outlook, with 48\% positive sentiment being the largest segment, followed by 42.9\% neutral sentiment, and only 7.21\% negative sentiment. This indicates that the overall communication style in the couples and group interactions tends to maintain a constructive and balanced emotional atmosphere, with very few instances of negative exchanges.
\\
Looking at the emotional aspects in the first chart, Desire (29.4\%) and Joy (23.9\%) emerge as the dominant emotions, collectively accounting for over half of the emotional expressions. This is followed by a notable segment of "None" (14.8\%) and "Interest" (9.03\%), suggesting that while interactions are generally emotionally engaged, there are also periods of neutral or emotionally reserved communication. The presence of other emotions like Anticipation, Surprise, and Excitement in smaller proportions indicates a rich diversity of emotional expression, though negative emotions like Anger remain minimal (3.04\%).
\\
The tone analysis in the second chart provides interesting insights into the communication style, with a Casual tone strongly dominating at 65.7\%, followed by a Formal tone at 17.2\%. This suggests that most interactions maintain a relaxed, comfortable atmosphere while still preserving some level of formality when needed. The presence of Neutral (7.89\%), Informal (5.26\%), and Playful (2.28\%) tones, with minimal Aggressive tone (1.67\%), indicates that the communication environment is generally conducive to open and comfortable interaction while maintaining appropriate boundaries and respect.

\begin{figure*}[ht]
  \centering
  \begin{minipage}[b]{0.48\linewidth}
    \includegraphics[width=\linewidth]{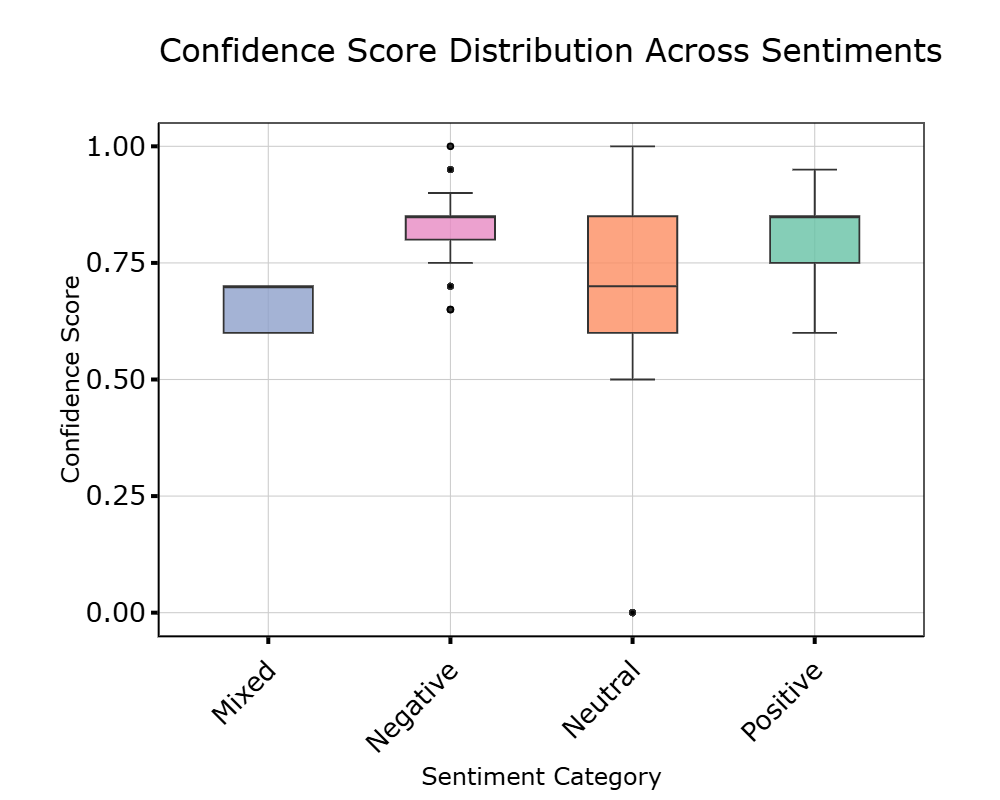}
    \caption*{(a) CCS}
  \end{minipage}
  \hfill
  \begin{minipage}[b]{0.48\linewidth}
    \includegraphics[width=\linewidth]{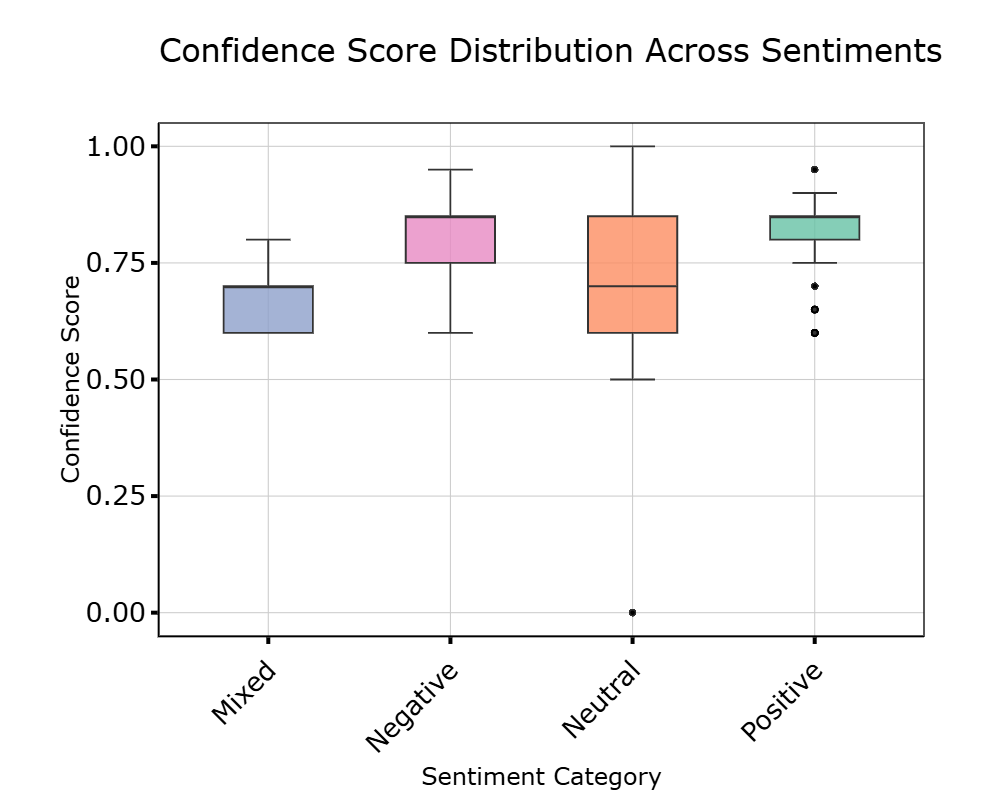}
    \caption*{(b) CGI}
  \end{minipage}

  \vspace{0.5cm}

  \begin{minipage}[b]{0.48\linewidth}
    \includegraphics[width=\linewidth]{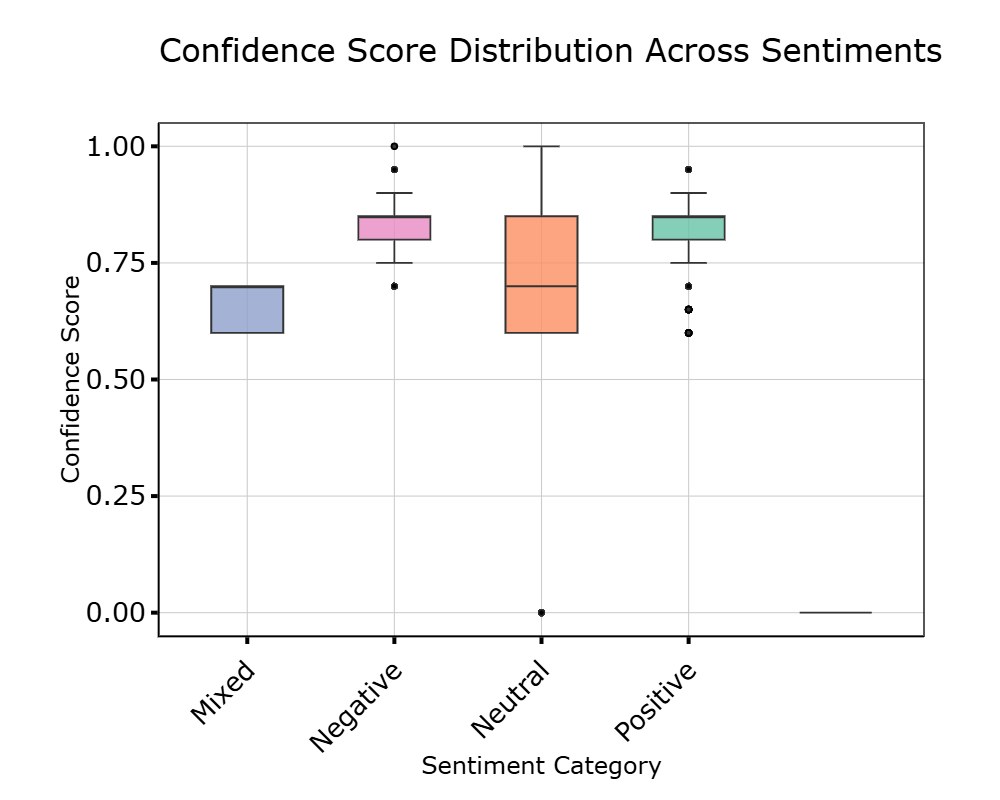}
    \caption*{(c) Ex}
  \end{minipage}
  \hfill
  \begin{minipage}[b]{0.48\linewidth}
    \includegraphics[width=\linewidth]{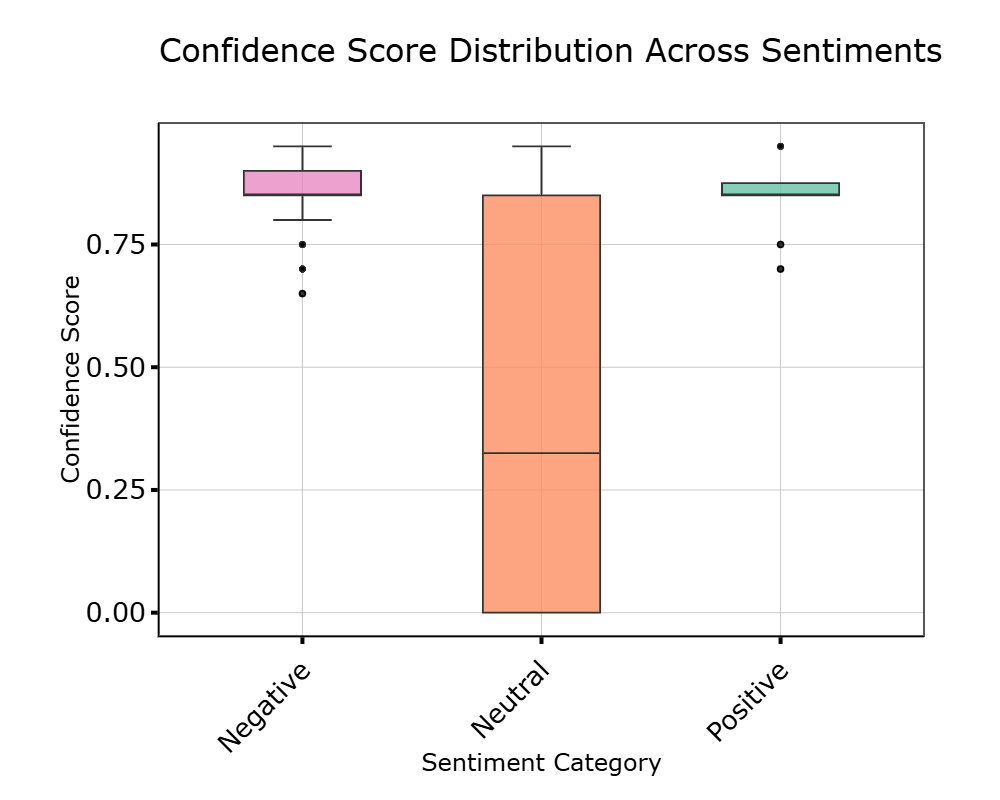}
    \caption*{(d) MF}
  \end{minipage}

  \vspace{0.5cm}

  \begin{minipage}[b]{0.48\linewidth}
    \includegraphics[width=\linewidth]{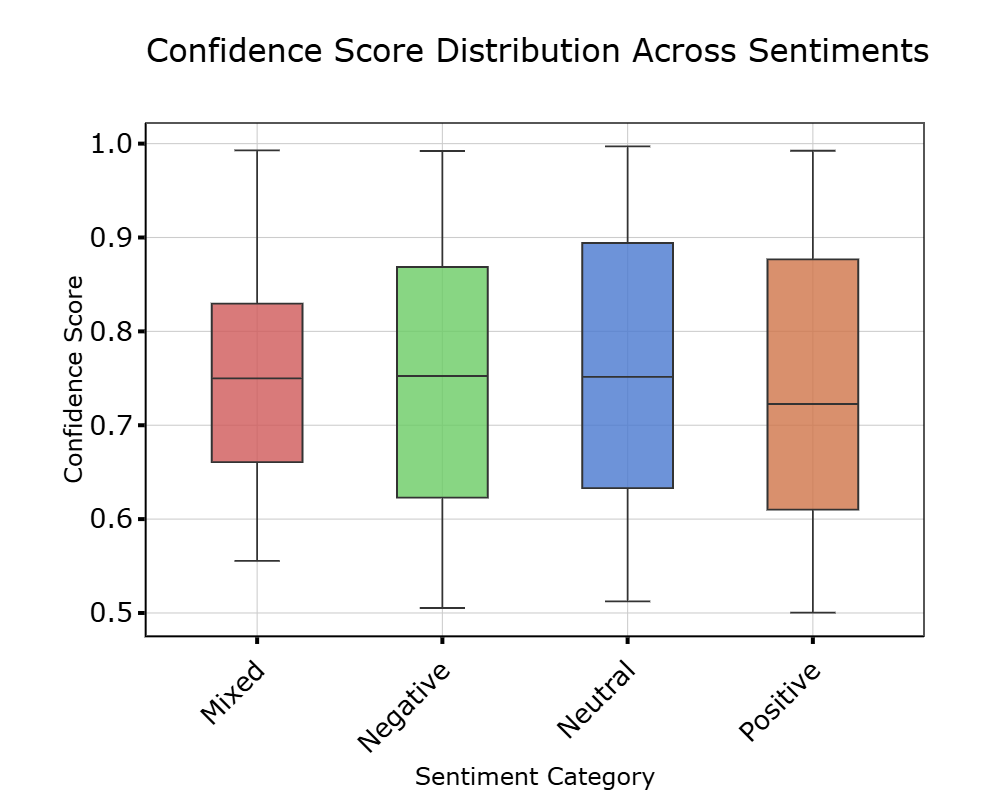}
    \caption*{(e) PS}
  \end{minipage}
  \hfill
  \begin{minipage}[b]{0.48\linewidth}
    \includegraphics[width=\linewidth]{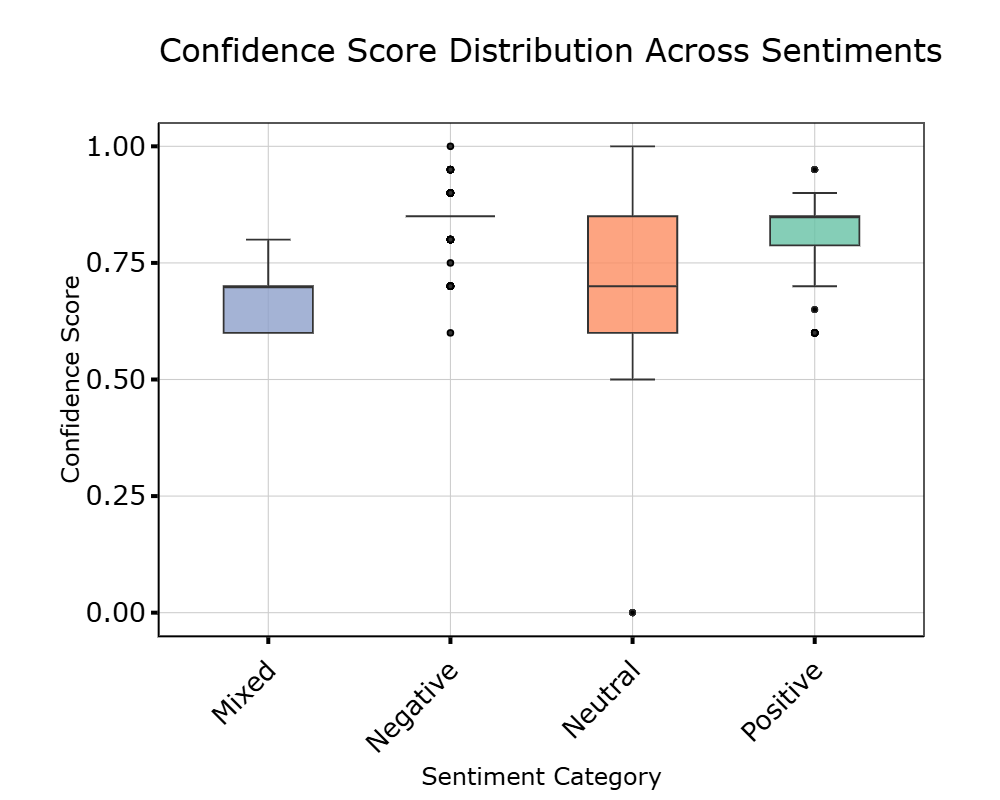}
    \caption*{(f) VS}
  \end{minipage}

  \caption{The Confidence Score Distribution by Sentiment: Box Plot Comparison of Positive, Neutral, Mixed, and Negative Categories}
  \label{figuregamma}
\end{figure*}

\clearpage

\textbf{ Confidence Score Distribution by Sentiment:}
From Figure.~\ref{figuregamma} the four sentiment categories (Positive, Neutral, Negative, Mixed). Overall, each box plot reveals moderate to high median confidence values, suggesting that the underlying model generally assigns sentiment labels with a notable degree of certainty. However, the presence of outliers and varying interquartile ranges in each subplot indicates that classification confidence can fluctuate depending on the specific context or linguistic cues present in the data.

A closer inspection of the individual subplots highlights subtle differences in how sentiments are classified. For instance, some classes, such as (d) MF, show relatively compact box plots for Positive sentiment, implying a more consistent level of confidence for positive classifications. In contrast, other classes (e.g., (a) CCS and (c) Ex) exhibit a broader spread for Neutral or Mixed sentiments, suggesting that the model occasionally encounters more ambiguity when distinguishing between emotionally neutral content and text that blends multiple affective tones. Negative sentiment typically shows slightly wider distributions, pointing to potential variability in how strongly negative cues are detected.

Collectively, these findings underscore a robust, yet context-sensitive classification process. While Positive sentiment often emerges with comparatively higher and more consistent confidence scores, Neutral, Mixed, and Negative categories reveal more diverse confidence intervals, reflecting the nuanced nature of human language and emotional expression. The recurring outliers across subplots further emphasize that certain instances may challenge the model’s ability to assign a definitive sentiment category. Overall, the distribution of confidence scores across these six classes illustrates a generally reliable classification framework, albeit one that must navigate the inherent complexities of the sentiment-laden text.

\subsection*{Sentimental Analysis using PLM (BERT)}

\begin{figure}[htbp]
    \centering
    \includegraphics[width=1\linewidth]{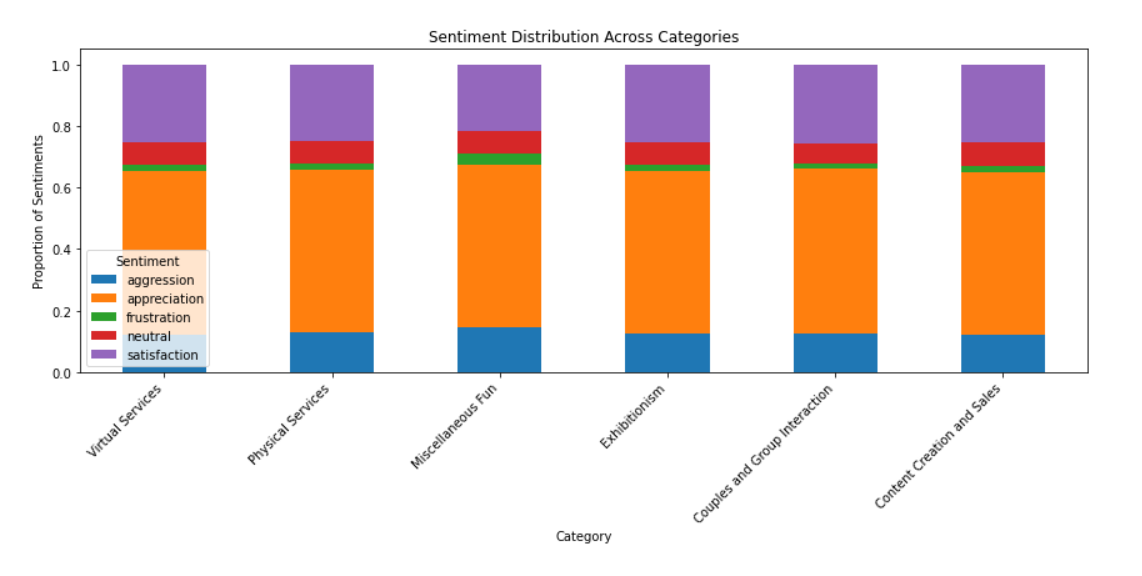}
    \caption{Sentiment Analysis using BERT Model}
    \label{fig:Bert}
\end{figure}

In reviewing Figure.~\ref{fig:Bert} describing the sentiment proportions across six categories a clear trend emerges. Positive sentiments such as “appreciation” and “satisfaction” occupy a substantial share in most categories, indicating that user feedback skews favorably. “Neutral” sentiment also appears consistently, though at varying levels, suggesting a notable fraction of content that neither leans strongly positive nor negative. In contrast, “aggression” and “frustration” are relatively lower, which might suggest that overtly negative expressions are less common in the overall dataset.

A closer look reveals subtle differences in sentiment composition among the categories. For example, Physical Services and Content Creations and Sales exhibit a higher prevalence of “appreciation,” pointing to more frequent expressions of gratitude or praise. Meanwhile, categories with more open-ended or interactive dynamics—such as Miscellaneous Fun—may see a slight rise in “frustration” or “aggression", suggesting sporadic instances of dissatisfaction. These observations are derived using a BERT-based model, which leverages contextual embeddings to classify text with a high degree of nuance. Consequently, the analysis highlights both the generally positive nature of user communications and the importance of contextual factors in shaping sentiment.

\clearpage

\section*{Appendix-3: Metadata of the Dataset}

\begin{figure}[htbp]
    \centering
    \includegraphics[width=\linewidth]{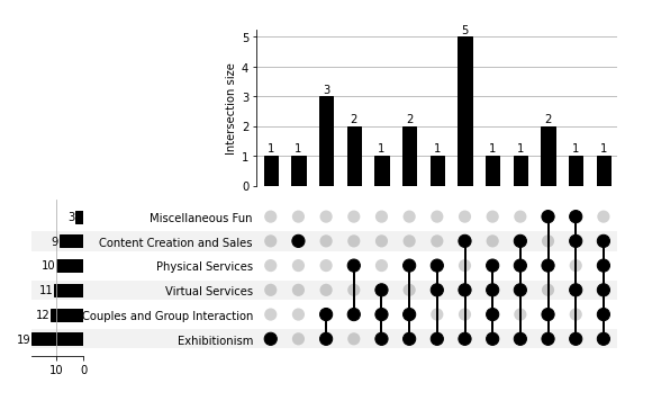}
    \caption{Total No. of Performers per Category and their Overlap}
    \label{fig:performer_overlap}
\end{figure}

Figure~\ref{fig:performer_overlap} provides a clear visual summary of how performers are distributed and overlap across multiple categories, including Miscellaneous Fun, Content Creation and Sales, Physical Services, Virtual Services, Couples and Group Interaction, and Exhibitionism. Each row corresponds to a category, and the black dots indicate shared performers among these categories. The bar chart at the top shows the number of performers participating in each specific combination of categories. Notably, “Exhibitionism” and “Couples and Group Interaction” are the most prevalent, as evidenced by the tallest bars, suggesting their high popularity or frequent reporting. Overall, the figure underscores that while many performers concentrate on a single category, a noteworthy subset engages in multiple overlapping areas, highlighting the importance of cross-category involvement.

Focusing on the intersection sizes depicted in the top bar chart, it is evident that most performers participate in only one or two categories. However, there is a distinct group of five performers who are active across a broader combination of categories, indicating a significant overlap in their services and interactions. Additionally, other smaller clusters—such as a group of three performers—reveal that although single-category involvement is common, a considerable minority diversifies their participation. These observations not only confirm that the majority of performers tend to specialize but also illuminate the multifaceted nature of the field, where cross-category engagement plays a crucial role in understanding performer behavior.

\clearpage

\section*{Appendix-4: Error Analysis}

We have performed error analysis and got the following results \textbf{(TP: True Positive, TN: True Negative, FP: False Positives, FN: False Negatives)}:


\begin{figure}[H]
    \centering
    \begin{minipage}{0.49\textwidth}
        \centering
        \includegraphics[width=\textwidth]{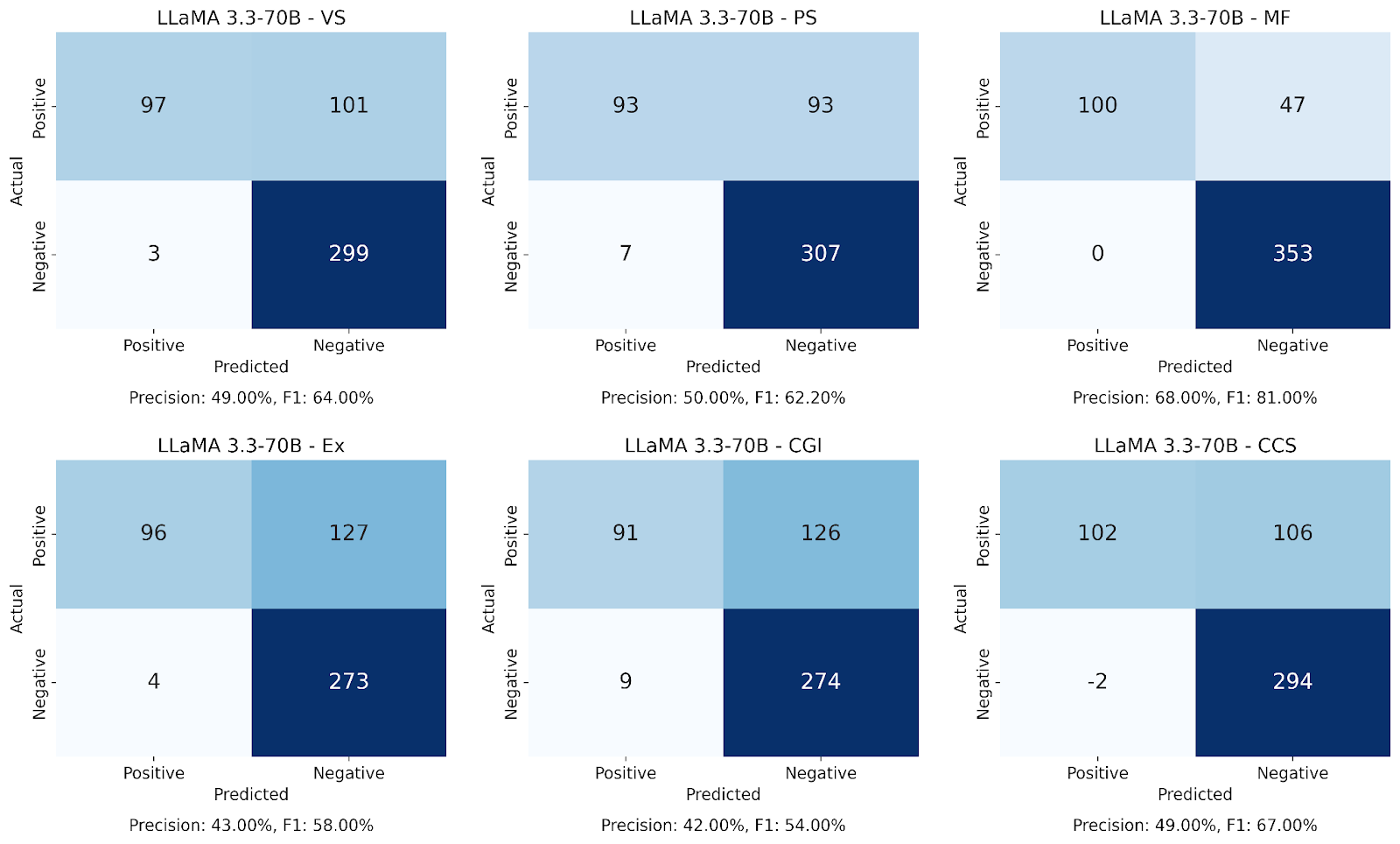}
        \caption{Confusion matrix for Llama 3.3 70B}
        \label{fig:cm1}
    \end{minipage}%
    \hfill
    \begin{minipage}{0.49\textwidth}
        \centering
        \includegraphics[width=\textwidth]{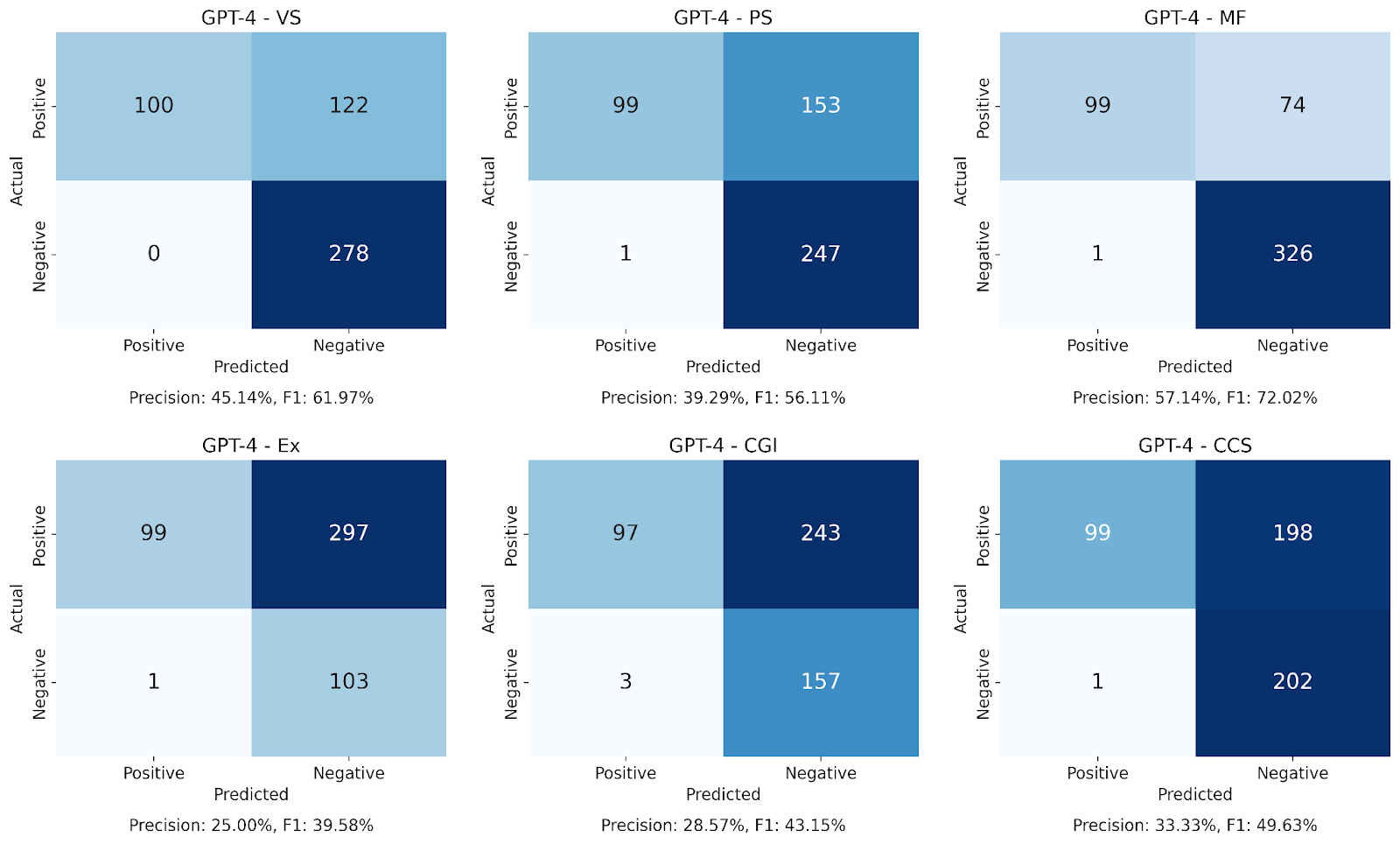}
        \caption{Confusion matrix for GPT 4}
        \label{fig:cm2}
    \end{minipage}
\end{figure}

\begin{figure}[H]
    \centering
    \begin{minipage}{0.49\textwidth}
        \centering
        \includegraphics[width=\textwidth]{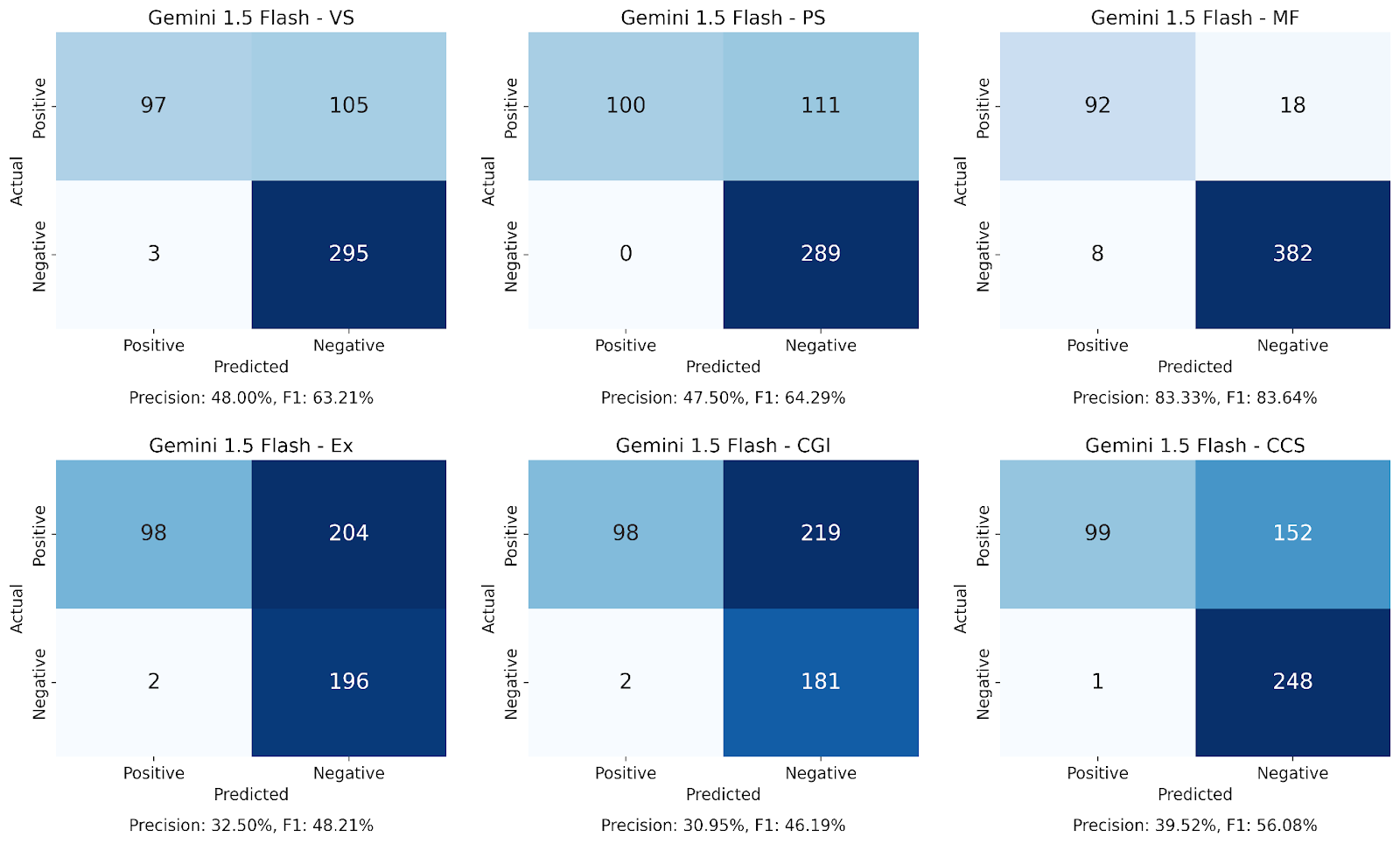}
        \caption{Confusion matrix for Gemini 1.5 Flash}
        \label{fig:cm3}
    \end{minipage}%
    \hfill
    \begin{minipage}{0.49\textwidth}
        \centering
        \includegraphics[width=\textwidth]{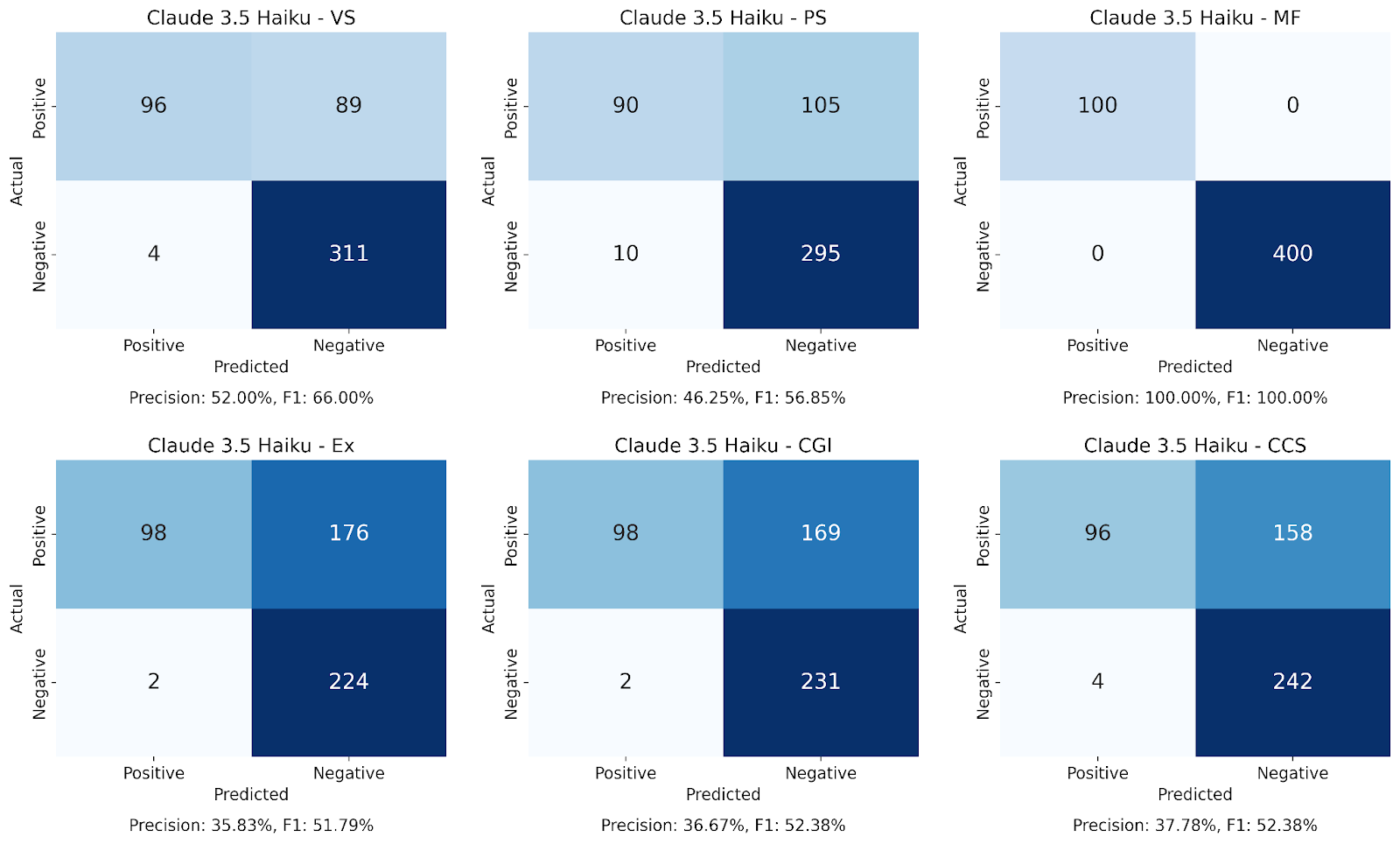}
        \caption{Confusion matrix for Claude 3.5 Haiku}
        \label{fig:cm4}
    \end{minipage}
\end{figure}

\begin{figure}[H]
    \centering
    \includegraphics[width=0.5\textwidth]{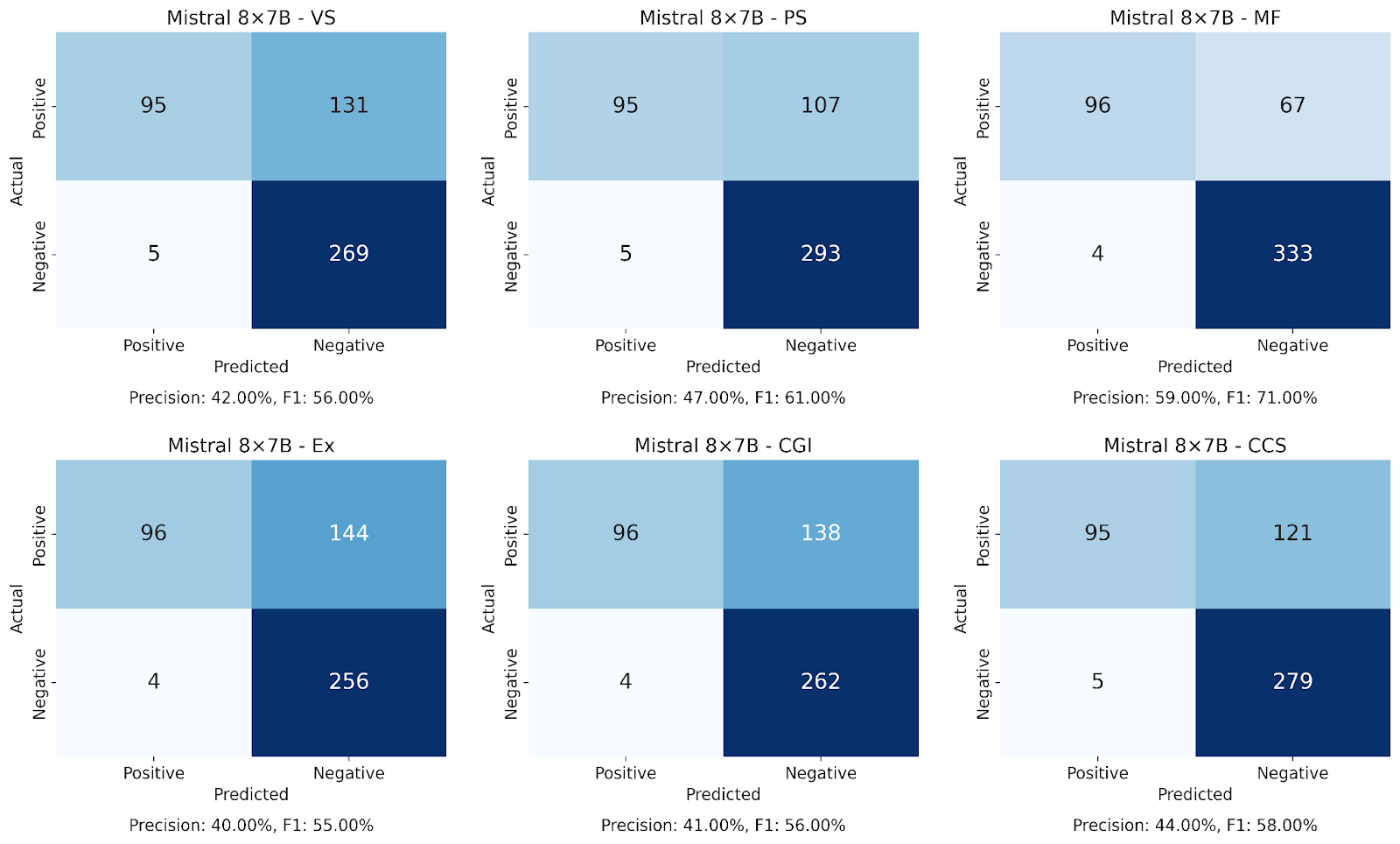}
    \caption{Confusion matrix for Mistral 8x7B}
    \label{fig:cm5}
\end{figure}

\subsection*{Class-by-Class Failure Analysis}

We have two types of data: the ground truth data and the LLM-predicted data. The ground truth data tells us which posts belong to which particular categories out of the six categories. We then compared this ground truth data with the predicted data from the LLM results. Our task is a multi-label classification task, i.e., each individual post may simultaneously belong to multiple categories. Therefore, evaluation metrics such as True Positive (TP), False Positive (FP), True Negative (TN), and False Negative (FN) are computed on a per-category basis, rather than per-post as a whole.

\subsubsection*{Definitions of Evaluation Metrics}

\begin{itemize}
    \item \textbf{True Positives (TP)}: Posts correctly identified as belonging to a specific category (present in both ground truth and predictions).
    \item \textbf{False Positives (FP)}: Posts incorrectly identified as belonging to a category (present in predictions but absent in ground truth).
    \item \textbf{False Negatives (FN)}: Posts belonging to a category that were missed by the model (present in ground truth but absent in predictions).
    \item \textbf{True Negatives (TN)}: Posts correctly identified as not belonging to a specific category (absent from both predictions and ground truth).
\end{itemize}

\textbf{Example:} Let's say Post\_A's ground truth classification is (Ex, VS, PS) and the predicted classification is (Ex, VS, CCS). Then:
\begin{itemize}
    \item TP: \{Ex, VS\} $\rightarrow$ 2 TPs
    \item FP: \{CCS\} $\rightarrow$ 1 FP
    \item FN: \{PS\} $\rightarrow$ 1 FN
    \item TN: All other labels (CGI, MF, etc.) not predicted and not in ground truth
\end{itemize}

\subsection*{Category-Wise Error Analysis}
\begin{enumerate}
    \item \textbf{VS (Virtual Services)}
    \begin{itemize}
        \item High FN for some models: Subtle language such as “online sessions” or “DM for details” sometimes goes undetected, increasing FN.
        \item FP: Certain LLMs mislabel purely conversational or flirtatious content as “VS” even when no explicit service is being offered.
    \end{itemize}
    \item \textbf{PS (Physical Services)}
    \begin{itemize}
    \item Lower FNs overall, suggesting that explicit phrases referencing physical meetups or location-based transactions are easier for the LLMs to pick up.
    \item Moderate FP: Posts hinting at offline gatherings for nonsexual contexts (e.g., “meet and greet,” “hangout”) occasionally get flagged as PS due to partial matching on keywords like “meet” or “services.”
    \end{itemize}
    \item \textbf{MF (Miscellaneous Fun)}
    \begin{itemize}
    \item Frequent Confusions: Many models confuse MF with either VS or PS because “miscellaneous fun” can overlap with playful or euphemistic language. This results in both FP (flagging other categories as MF) and FN (failing to catch truly offbeat or playful sexual transactions).
    \item The inconsistent usage of slang is a leading cause of labeling errors here.
    \end{itemize}
    \item \textbf{Ex (Exhibitionism)}
    \begin{itemize}
    \item High FN: Models often fail to classify content as Ex if it doesn’t contain explicit words like “public,” “show,” or “watch,” thus missing subtle references (e.g., partial mention of webcam exhibition).
    \item FP: Overzealous classification of normal adult posts as Ex if they contain words like “display,” “pics,” or “look at me.”
    \end{itemize}
    \item \textbf{CGI (Couples and Group Interactions)}
    \begin{itemize}
    \item Contextual Mistakes: Many false negatives occur when more than two people are mentioned, but the nature of the post isn’t strictly a couples/group sexual context (e.g., “We’re looking for new friends” can be mistaken for group sexual activity).
    \item FPs are rarer but happen when the LLM sees terms like “we” or “our” and incorrectly jumps to CGI.
    \end{itemize}
    \item \textbf{CCS (Content Creation and Services)}
    \begin{itemize}
    \item High Overlap with VS: Offers for camming, custom videos, or phone services sometimes get mislabeled as plain Virtual Services. This confusion causes FP under CCS and FN under VS—or vice versa—depending on the model’s bias.
    \item Industry Slang: Phrases like “collab,” “premium account,” or “pay-per-view” are sometimes overlooked, causing FNs in genuine CCS posts.
    \end{itemize}  
\end{enumerate}






\subsection*{Model-Specific Observations}

\begin{itemize}
    \item \textbf{GPT-4} gives high precision overall, rarely flagging benign content as NSFW, but it often misses coded or subtle references (especially in Ex and MF).
    \item \textbf{Claude} demonstrates balanced detection with moderate FP and FN but struggles to differentiate borderline PS from MF in cases mixing mild in-person hints with casual context.
    \item \textbf{Llama} maintains the lowest FP in Ex/CGI by not over-flagging posts referencing “we” or “together,” yet it exhibits the highest FN in precisely those categories, indicating a conservative approach that overlooks subtle group/exhibition content.
    \item \textbf{Mistral}, with relatively strong recall for VS and CCS due to recognizing industry terms (e.g., “OnlyFans,” “cam sessions”), generates more false positives for PS when vague in-person language or “arrangements” appear.
    \item \textbf{Gemini} excels at identifying partial or suggestive references to exhibitionist scenarios (Ex), but overestimates group activity (CGI), leading to inflated FPs, and at times incorrectly labels broader adult entertainment topics as CCS, thereby confusing them with PS or MF.
\end{itemize}

\subsection*{Key Observations}

\begin{itemize}
    \item \textbf{Coded or slang language} remains a major source of false negatives (FNs) across GPT-4, Claude, and Llama, as euphemisms, abbreviations, or local jargon go undetected.
    \item \textbf{Mistral} handles digital service terms slightly better but overestimates physical services references.
    \item \textbf{Gemini} identifies subtle Exhibitionism cues yet often conflates them with broader adult content, causing false positives in CGI or CCS.
    \item All models grapple with \textbf{contextual ambiguity} when posts vaguely reference “arrangements” or “fun” without explicit mention of money or sex. GPT-4 and Claude sometimes over-flag borderline content, Llama under-flags it, and Mistral and Gemini show mixed outcomes depending on the transaction hints.
    \item \textbf{Class overlaps} also pose challenges: Mistral and Llama frequently confuse VS and CCS if a post mentions an online platform but doesn’t explicitly clarify “content creation,” and Gemini over-detects group aspects in Ex, while GPT-4 fails to catch borderline public exhibition references.
\end{itemize}

\clearpage

\section*{Appendix-5: Inter Annotator Agreement scores}

The labelling process was conducted by a team of three annotators: two male and one female. The inclusion of a female annotator was intentional to mitigate potential gender bias in the dataset, ensuring that sexual services offered by both men and women were adequately represented. Each annotator was responsible for reviewing the activities of users active within the selected subreddits and analyzing the services they offered. Based on this analysis, the content was categorized and labeled according to the type of services provided. The following categories were established:

\begin{itemize}
    \item Content Creation and Sales (CCS): Activities related to producing and selling explicit material.
    \item Couples and Group Interactions (CGI): Services involving multiple participants.
    \item Exhibition (Ex): Public or performative displays of sexual content.
    \item Miscellaneous Fun (MF): Uncategorized or recreational sexual activities.
    \item Physical Services (PS): In-person sexual services.
    \item Virtual Services (VS): Online or remote sexual interactions.
\end{itemize}

These categories provided a structured framework for analyzing the range of services present within the dataset.

We have evaluated inter annotator agreement scores. We have also calculated parameters like Cohen and Fleiss Kappa scores. The table representing all these scores is given below:

\begin{table}[h]
\small
\centering
\begin{tabular}{cccc}
\hline
Annotator Pair & Krippendorff ($K$) & Cohen ($C$) & Fleiss ($F$) \\
\hline
(1,2) & 0.6633 & 0.554 & --- \\
(1,3) & 0.7470 & 0.681 & --- \\
(2,3) & 0.6783 & 0.740 & --- \\
(1,2,3) & 0.6963 & --- & 0.608 \\
\hline
\end{tabular}
\caption{Inter-Annotator Agreement Scores}
\end{table}

For the evaluation of the IAA scores we have presented the values that are annotated on the entire dataset, three annotators (denoted as 1, 2, and 3) were assessed for agreement across their annotations. The have the interrater reliability scores using pairwise Krippendorff scores were calculated as follows: $K(1,2) = 0.6633, K(1,3) = 0.7470,$ and $K(2,3) = 0.6783$. For the collective agreement among all three annotators, $K(1,2,3) = 0.6963$.

We have also presented Cohen's kappa score $C(1,2) = 0.554, C(2,3) = 0.681$ and $C(1.3) = 0.74$. We have also evaluated the Fleiss kappa $F(1,2,3)= 0.608$. Collectively, these values indicate a substantial level of inter-annotator agreement and interrater reliability, demonstrating consistency and reliability in the annotation scores across annotators.

\subsection*{Procedure Involving Data Collection and Construction}
\begin{enumerate}
    \item \textbf{Data Collection}
    The data for this study was collected from three specific subreddits identified as primary hubs for discussions related to sexual services. Data extraction was performed using the Reddit API, facilitated by the PRAW (Python Reddit API Wrapper) library, which enabled the retrieval of both posts and comments from these subreddits.
    \item \textbf{Data Cleaning}
    The initial dataset underwent a cleaning process to eliminate irrelevant or extraneous content. Posts and comments deemed non-substantive, such as greetings (e.g., "Hi," "Hello!"), were removed to ensure the dataset focused solely on meaningful exchanges related to the research topic.
    \item \textbf{Data Preprocessing}
    To protect the anonymity of individuals involved, several preprocessing steps were implemented. Posts containing visible faces were excluded from the dataset, as most posts naturally blurred such identifying features. Additionally, all usernames and Reddit IDs were stripped from the data, retaining only the content of the posts and comments for analysis.
    
\end{enumerate}

\clearpage

\section*{Appendix-6:  Statistical significance of dataset}

We analyze several features associated with posts, including post file size, post title length, high media proportion, and image count. For each category, the mean number of images per post is computed. Posts with an image count exceeding this mean are classified as high media posts, and the proportion of such posts within each category is used as a derived metric. Among the considered features, post file size is treated as the primary metric, as it encapsulates multiple content dimensions of a post. The following tables present the computed t-statistics and p-values for these features across all pairwise category comparisons.

\subsection*{Metric: Post File Size}
\begin{table}[H]
\small
\centering
\begin{tabular}{ccccc}
\hline
\textbf{Category 1} & \textbf{Category 2} & \textbf{Mean 1} & \textbf{Mean 2} & \textbf{P-value} \\
\hline
VS & PS & 692.782 & 1073.87 & 1.32074e-06 \\
VS & MF & 692.782 & 61.7645 & 9.7284e-36 \\
VS & Ex & 692.782 & 1288.55 & 3.39456e-14 \\
VS & CGI & 692.782 & 2322.52 & 3.27258e-35 \\
VS & CCS & 692.782 & 1309.01 & 6.71561e-12 \\
PS & MF & 1073.87 & 61.7645 & 3.17274e-48 \\
PS & Ex & 1073.87 & 1288.55 & 0.0176972 \\
PS & CGI & 1073.87 & 2322.52 & 1.09732e-19 \\
PS & CCS & 1073.87 & 1309.01 & 0.0190976 \\
MF & Ex & 61.7645 & 1288.55 & 3.7581e-71 \\
MF & CGI & 61.7645 & 2322.52 & 4.82063e-68 \\
MF & CCS & 61.7645 & 1309.01 & 1.06532e-52 \\
Ex & CGI & 1288.55 & 2322.52 & 4.08788e-14 \\
CGI & CCS & 2322.52 & 1309.01 & 1.52351e-12 \\
\hline
\end{tabular}
\caption{T-statistics and p-values for Post File Size}
\end{table}

\subsection*{Metric: Image Count Post-wise}
\begin{table}[H]
\small
\centering
\begin{tabular}{ccccc}
\hline
\textbf{Category 1} & \textbf{Category 2} & \textbf{Mean 1} & \textbf{Mean 2} & \textbf{P-value} \\
\hline
VS & PS & 0.417557 & 0.426999 & 0.00623566 \\
VS & MF & 0.417557 & 0.0857143 & 7.65742e-21 \\
VS & Ex & 0.417557 & 0.4596 & 0.0149964 \\
VS & CGI & 0.417557 & 0.555757 & 1.34459e-11 \\
VS & CCS & 0.417557 & 0.490145 & 8.5577e-05 \\
PS & MF & 0.426999 & 0.0857143 & 9.60819e-22 \\
PS & Ex & 0.426999 & 0.4596 & 0.0555793 \\
PS & CGI & 0.426999 & 0.555757 & 1.86928e-10 \\
PS & CCS & 0.426999 & 0.490145 & 0.000535663 \\
MF & Ex & 0.0857143 & 0.4596 & 6.08182e-25 \\
MF & CGI & 0.0857143 & 0.555757 & 2.98606e-33 \\
MF & CCS & 0.0857143 & 0.490145 & 1.60089e-27 \\
Ex & CGI & 0.4596 & 0.555757 & 1.51046e-07 \\
CGI & CCS & 0.555757 & 0.490145 & 0.000717709 \\
\hline
\end{tabular}
\caption{T-statistics and p-values for Image Count}
\end{table}

\subsection*{Metric: High Media Proportion Count}
\begin{table}[H]
\small
\centering
\begin{tabular}{ccccc}
\hline
\textbf{Category 1} & \textbf{Category 2} & \textbf{Proportion 1} & \textbf{Proportion 2} & \textbf{P-value} \\
\hline
VS & PS & 41.5267 & 42.6266 & 0.564769 \\
VS & MF & 41.5267 & 8.57143 & 2.75455e-11 \\
VS & Ex & 41.5267 & 45.8297 & 0.0123093 \\
VS & CGI & 41.5267 & 55.5757 & 5.92104e-12 \\
VS & CCS & 41.5267 & 46.9398 & 0.00236237 \\
PS & MF & 42.6266 & 8.57143 & 6.96976e-12 \\
PS & Ex & 42.6266 & 45.8297 & 0.00059388 \\
PS & CGI & 42.6266 & 55.5757 & 1.57679e-10 \\
PS & CCS & 42.6266 & 46.9398 & 0.0143173 \\
MF & Ex & 8.57143 & 45.8297 & 5.57332e-14 \\
MF & CGI & 8.57143 & 55.5757 & 0 \\
MF & CCS & 8.57143 & 46.9398 & 1.39888e-14 \\
Ex & CGI & 45.8297 & 55.5757 & 1.01586e-07 \\
CGI & CCS & 55.5757 & 46.9398 & 4.76368e-06 \\
\hline
\end{tabular}
\caption{T-statistics and p-values for High Media Proportion Count}
\end{table}

\subsection*{Metric: Post Title Length}
\begin{table}[H]
\small
\centering
\begin{tabular}{ccccc}
\hline
\textbf{Category 1} & \textbf{Category 2} & \textbf{Mean 1} & \textbf{Mean 2} & \textbf{P-value} \\
\hline
VS & PS & 46.0916 & 40.8694 & 0.00670334 \\
VS & MF & 46.0916 & 88.2286 & 0.000665298 \\
VS & Ex & 46.0916 & 45.1703 & 0.599449 \\
VS & CGI & 46.0916 & 60.3336 & 0.0688431 \\
VS & CCS & 46.0916 & 45.4077 & 0.695139 \\
PS & MF & 40.8694 & 88.2286 & 0.000143854 \\
PS & Ex & 40.8694 & 45.1703 & 0.00111554 \\
PS & CGI & 40.8694 & 60.3336 & 0.00128002 \\
PS & CCS & 40.8694 & 45.4077 & 0.00709253 \\
MF & Ex & 88.2286 & 45.1703 & 0.000497275 \\
MF & CGI & 88.2286 & 60.3336 & 0.0508422 \\
MF & CCS & 88.2286 & 45.4077 & 0.000531429 \\
Ex & CGI & 45.1703 & 60.3336 & 0.0511451 \\
CGI & CCS & 60.3336 & 45.4077 & 0.0548228 \\
\hline
\end{tabular}
\caption{T-statistics and p-values for Post Title Length}
\end{table}

In the tables presented, Mean 1 and Mean 2 denote the average values of the respective metrics computed across the compared categories. A consistent pattern of statistically significant differences is observed in comparisons involving the MF category relative to others.

\textbf{Note:} The post title length metric demonstrates limited utility for statistical significance testing. Due to the brevity and semantic sparsity of many titles, this feature offers reduced discriminative power, thereby compromising the reliability and interpretability of p-value estimates based on this variable.

Given that the vast majority of p-values obtained are substantially small, the application of standard multiple-testing correction procedures (e.g., Bonferroni or Benjamini-Hochberg) would still yield corrected p-values that remain well below conventional significance thresholds. Therefore, the implementation of such corrections does not materially affect our overall conclusions regarding the statistical significance of inter-category differences.

\clearpage

\section*{Appendix-7: Ensemble Methods}

The ensemble results reveal several key patterns. Triadic combinations (e.g., G-C-M, G-L-M) tend to perform better than dyadic ones (e.g., G-L, L-M) across several metrics, particularly in accuracy and F1 scores for categories like VS and CGI. Notably, the G-C-M ensemble achieves the highest overall accuracy (63.24\%) and the lowest JSD (0.125), indicating strong alignment with ground truth distributions. Interestingly, while exact match precision (Ex) remains highest for individual models (100\%), its F1 drops sharply in combinations involving L and C, suggesting trade-offs between precision and coverage. MSE and MAE values remain relatively stable across configurations, showing marginal gains with trios. Overall, ensemble aggregation—especially among diverse models—appears to moderately improve performance consistency and robustness.

\begin{table*}[hbt]
\small
\centering
\resizebox{1.0\textwidth}{!}{%
\begin{tabular}{lccccccccccc}
\hline
\textbf{Metric} & \textbf{G-L} & \textbf{G-C} & \textbf{G-M} & \textbf{L-C} & \textbf{L-M} & \textbf{C-M} & \textbf{G-L-C} & \textbf{G-L-M} & \textbf{G-C-M} & \textbf{L-C-M} \\
\hline
Pre (\%) VS & 70.59 & 70.59 & 70.59 & 61.33 & 61.33 & 65.66 & 67.89 & 72.12 & 74.23 & 63.40 \\
F1 (\%) VS & 15.89 & 15.89 & 15.89 & 22.44 & 22.44 & 26.08 & 19.00 & 19.38 & 18.77 & 23.57 \\
Pre (\%) PS & 53.11 & 53.11 & 53.11 & 52.59 & 52.59 & 52.02 & 52.88 & 54.51 & 54.01 & 52.13 \\
F1 (\%) PS & 47.42 & 47.42 & 47.42 & 51.71 & 51.71 & 51.87 & 51.40 & 47.15 & 47.51 & 50.52 \\
Pre (\%) Ex & 100.00 & 100.00 & 100.00 & 96.83 & 96.83 & 98.03 & 98.63 & 98.87 & 98.98 & 97.69 \\
F1 (\%) Ex & 40.10 & 40.10 & 40.10 & 17.11 & 17.11 & 20.52 & 28.44 & 33.59 & 36.61 & 17.76 \\
Pre (\%) CGI & 85.00 & 85.00 & 85.00 & 86.18 & 86.18 & 86.86 & 85.61 & 86.54 & 86.01 & 87.95 \\
F1 (\%) CGI & 17.19 & 17.19 & 17.19 & 25.17 & 25.17 & 23.20 & 22.14 & 25.84 & 23.84 & 27.68 \\
Pre (\%) CCS & 58.47 & 58.47 & 58.47 & 60.00 & 60.00 & 59.26 & 61.96 & 55.64 & 56.78 & 59.78 \\
F1 (\%) CCS & 19.03 & 19.03 & 19.03 & 14.74 & 14.74 & 5.05 & 16.31 & 20.00 & 18.48 & 15.74 \\
MSE & 0.464 & 0.464 & 0.464 & 0.486 & 0.486 & 0.475 & 0.471 & 0.465 & 0.461 & 0.482 \\
MAE & 0.464 & 0.464 & 0.464 & 0.486 & 0.486 & 0.475 & 0.471 & 0.465 & 0.461 & 0.482 \\
JSD & 0.128 & 0.128 & 0.128 & 0.142 & 0.142 & 0.144 & 0.134 & 0.126 & 0.125 & 0.139 \\
Acc (\%) & 62.43 & 62.43 & 62.43 & 55.88 & 55.88 & 56.10 & 60.19 & 62.28 & 63.24 & 57.07 \\
\hline
\end{tabular}
}
\caption{Evaluation Results of aggregation of LLMs (ensemble methods). Here, G\(\rightarrow\) Gemini 1.5 Flash, L\(\rightarrow\) LlaMA 3.3-70B-Instruct, M\(\rightarrow\) Mistral 8\(\times\)7B, Q\(\rightarrow\) Qwen 2.5 Turbo, C\(\rightarrow\) Claude 3.5 Haiku.}
\end{table*}

\noindent \textbf{Note:} Six evaluation metrics were employed to assess model performance: Precision, F1 Score, Mean Squared Error (MSE), Mean Absolute Error (MAE), Jensen-Shannon Divergence (JSD), and Accuracy. The evaluation methodology was consistent with that used for the individual large language models (LLMs) described earlier in the paper, wherein the predicted outputs of the model were systematically compared against the ground truth annotations.

\section*{Appendix-8: Ablation studies on feature importance}

The evaluation yielded several key observations regarding the contribution of different feature sets to model performance.

\textbf{Exclusion of Emotion Features:} The model maintains near-perfect performance (approximately 0.99) across all major evaluation metrics, including Accuracy, Precision, Recall, and F1 Score, even when emotion-related features are removed. This suggests that these features may be redundant or potentially introduce noise, as their exclusion does not degrade the model’s predictive capabilities.

\textbf{Use of Sentiment, Emotion, or Tone Features in Isolation:} When the model is trained using only sentiment features, only emotion features, or only tone features, performance drops considerably. Metric values in these configurations range between approximately 0.07 and 0.14, indicating that while each feature set provides some predictive signal, none is sufficient on its own to support robust classification.

\textbf{Exclusion of Comment Features:} The model exhibits extremely poor performance—around 0.07—when comment-related features are removed. This sharp decline underscores the central role these features play in the model’s functioning. Their absence leads to a near-total failure in classification, indicating that comment features are foundational to the model’s success.

\textbf{Exclusion of Sentiment, Tone, and Metadata Features:} The removal of any of these feature sets results in performance approaching zero. This clearly demonstrates that each type of feature is essential. Sentiment features capture affective nuances, tone features contribute to contextual interpretation, and metadata features offer critical structural insights that support classification accuracy.

\begin{table}[H]
\small
\centering
\begin{tabular}{lcccc}
\hline
\textbf{Feature Set} & \textbf{Accuracy} & \textbf{Precision} & \textbf{Recall} & \textbf{F1 Score} \\
\hline
No Sentiment Features & 0.03 & 0.04 & 0.03 & 0.03 \\
No Emotion Features & 0.99 & 0.98 & 0.99 & 0.99 \\
No Tone Features & 0.02 & 0.03 & 0.02 & 0.01 \\
No Comment Features & 0.07 & 0.06 & 0.09 & 0.06 \\
No Metadata Features & 0.03 & 0.03 & 0.02 & 0.03 \\
Sentiment Features Only & 0.14 & 0.12 & 0.13 & 0.11 \\
Emotion Features Only & 0.07 & 0.08 & 0.07 & 0.07 \\
Tone Features Only & 0.09 & 0.09 & 0.08 & 0.09 \\
\hline
\end{tabular}
\caption{Table representing ablation study on feature importance.}
\end{table}

These results collectively indicate that ReddiX-NET depends on a well-balanced combination of features for effective performance. In particular, comment features and LLM-derived features—namely sentiment and tone—are indispensable. Their contribution is vital for achieving high classification accuracy and maintaining model robustness.

\clearpage

\section*{Appendix-9: Discussion and Policy Implications}

This research introduces \textbf{ReddiX-NET}, a novel dataset created to address a significant gap in online content moderation: the detection and regulation of online sexual services. Traditional content moderation tools primarily focus on explicit imagery using NSFW filters, but they struggle with the more nuanced challenge of identifying the \textit{solicitation} of sexual services online.

\vspace{0.5em}
\noindent\textbf{Motivation:} Current AI moderation tools are limited in their ability to distinguish between legal adult content and illegal solicitation of sexual services. The ReddiX-NET dataset specifically targets this distinction, which is essential for applications such as:
\begin{itemize}
    \item Enhancing the capability of content moderation systems.
    \item Assisting law enforcement in tracking online prostitution networks.
    \item Supporting psychological research on the impact of such content.
\end{itemize}

\subsection*{Psychological and Emotional Impact}

\begin{itemize}
    \item \textbf{Mental Health Relevance:} One key question addressed by the analysis is: \textit{How do different online sexual services affect users psychologically?} This is especially relevant for mental health professionals and behavior analysts.
    
    \item \textbf{Ethical and Emotional Ramifications:} The psychological impact data underscores the significance of this research—not only from a legal or ethical standpoint but also due to the emotional consequences such services have on users. These services can significantly alter user behavior, reflecting why restrictions were originally placed on such interactions.

    \item \textbf{Emotional-Tone Correlation:} Cross-correlation analysis reveals that specific emotional patterns tend to precede problematic interactions. This helps in drawing connections between emotion, tone, and user behavior, which can be instrumental in early detection.
\end{itemize}

\subsection*{Temporal Analysis for Intervention Timing}

    The temporal analysis, when combined with sentiment data, identifies not only when users are most active, but also when they exhibit the most concerning emotional patterns. This insight is valuable for determining the time windows during which interventional resources should be deployed, potentially helping platforms enforce policies and curb exploitation more effectively.

\subsection*{Implications for Policy Development}

\textbf{Key Policy Questions Addressed:}
\begin{itemize}
    \item \textbf{Which categories require different moderation approaches?}
    
        Categories such as VS (Virtual Services) and Ex (Exhibitionism), which evoke emotions like joy and desire, might require softer intervention strategies. In contrast, PS (Physical Services)—linked with higher degrees of lust and disgust—may necessitate stricter, more targeted moderation techniques.

    \item \textbf{When is additional user protection necessary?}
    
        Posts exhibiting higher emotional dependency metrics (e.g., desperation, obsession, emotional manipulation) may call for escalated protective mechanisms, such as automated warnings or human moderation.

    \item \textbf{How can platforms distinguish between legal adult content and harmful interactions?}
   
        The detailed breakdowns of sentiment and tone provide linguistic markers that differentiate consensual content from exploitative or illegal solicitation. This information can be integrated into AI-based classifiers to refine moderation filters beyond surface-level keyword detection.
    
\end{itemize}

\clearpage

\section*{Appendix-10: More related works}

Previous research in this domain relied primarily on conventional machine learning and pre-trained language model (PLM)-based approaches:
  
\cite{ibanez2016trafficking} Utilized network analysis and content matching to identify trafficking indicators, focusing on traditional NLP techniques rather than generative AI models. Limited to pattern recognition without the contextual understanding LLMs provide.

\cite{diaz2020natural} Developed classifiers for trafficking detection by training on illegal business data using standard classification methods. Did not leverage the advanced language understanding capabilities of LLMs.

\cite{wang2020sex} Employed ordinal regression neural networks, which, while sophisticated, lacked the comprehensive pretrained knowledge and contextual understanding inherent to modern LLMs.

Traditional NSFW Filters:
Typically rely on keyword matching and image recognition systems. Fail to capture the complex, evolving language patterns used to evade detection.

\textbf{Our Approach}:

Our approach fundamentally differs by incorporating state-of-the-art LLMs (GPT-4, LLaMA 3.3-70B, Gemini 1.5 Flash, Mistral 8×7B, Claude 3.5 Haiku) alongside BERT-based models, creating several advantages over these prior works:

\textbf{Contextual Understanding}: 
  - While prior work focused on explicit keywords or patterns, our LLM-based approach captures implicit meaning and coded language—a critical capability for detecting sophisticated evasion tactics.

\textbf{Multi-dimensional Analysis}:  
  - We uniquely combine LLMs for classification tasks with BERT-based models for sentiment analysis, creating a comprehensive analytical framework that assesses both content categorization and emotional dimensions.

\textbf{Zero-shot Learning Capabilities}:  
  - Unlike previous approaches requiring extensive labeled data, our LLM integration enables effective classification with minimal supervised training, adapting to emerging patterns more efficiently.

\textbf{Nuanced Category Detection}:  
  - Prior works primarily focused on binary classification (illegal/legal), whereas our approach leverages LLMs' advanced discriminative capabilities to distinguish between six distinct service categories with greater precision.

\textbf{Deeper Linguistic Processing}:  
  - Our methodology employs LLMs to analyze tone, sentiment, and emotional dependencies—aspects largely overlooked in previous research that relied on more superficial text characteristics.

The integration of LLMs in ReddiX-NET represents a methodological leap forward, enabling detection of previously unidentifiable patterns and providing richer analytical insights than traditional approaches. This hybrid PLM-LLM framework sets a new benchmark for content moderation systems, especially in domains where language is deliberately obfuscated to evade detection.

Advantages of the 6-Category Classification Approach:
Traditional NSFW detection and content moderation systems typically employ binary classification (appropriate/inappropriate or legal/illegal), which severely limits their effectiveness when dealing with the complex landscape of online sexual services. The 6-category classification approach in ReddiX-NET (Virtual Services, Physical Services, Miscellaneous Fun, Exhibitionism, Couples and Group Interactions, and Content Creation and Sales) provides significantly enhanced insights and capabilities:

- Reduced False Positives:
  - Binary systems frequently miscategorize legal adult content as solicitation, leading to over-moderation. Multi-class classification reduces these errors by properly distinguishing between categories.

- Enhanced Cross-Category Learning: 
  - The model learns discriminative features between similar categories (e.g., Virtual vs. Physical services), improving representation learning for all categories.
 
 We will compare two of those paper’s results with our results and have a detailed analysis of that.

\cite{diaz2020natural} The primary objective is to automatically detect illicit activity in business reviews, specifically targeting massage businesses. The paper aims to develop a classifier that can differentiate between legitimate and illicit Yelp reviews, particularly those associated with businesses listed on Rubmaps—a platform known for reviews that often reference illicit services. Below is the table summarizing its results. (This result was in graphs; we had to extrapolate it to a table, so the values are approximate).

\begin{table}[H]
\small
\centering
\caption{(\cite{diaz2020natural}) - Model Performance under Various Sparsity Levels}
\begin{tabular}{lcccc}
\hline
\textbf{Model} & \textbf{94\% Sparsity} & \textbf{95\% Sparsity} & \textbf{97\% Sparsity} & \textbf{No Sparsity Removal} \\
\hline
NB & 0.61 & 0.61 & 0.62 & 0.63 \\
NN & 0.65 & 0.66 & 0.66 & 0.67 \\
RF & 0.74 & 0.75 & 0.76 & 0.76 \\
SVM & 0.70 & 0.72 & 0.74 & 0.75 \\
\hline
\end{tabular}
\end{table}

\noindent Here, NB: Naive Bayes, NN: Neural Networks, RF: Random Forest, and SVM: Support Vector Machine respectively.

\cite{wang2020sex} The paper is not directly related to NSFW detection but is a method paper which proposes a strong method to predict labels using text reviews (similar to comment and post classification). The paper aims to predict ordinal labels (e.g., likelihood scores from 1 to 7) using text reviews. It addresses the unique challenge of ordered classes, which standard classifiers often ignore. To do this, the authors propose a novel model ORNN (Ordinal Regression Neural Network).

\begin{table}[H]
\small
\centering
\caption{Wang et al. (2020) - Model Evaluation Results}
\begin{tabular}{lcccc}
\hline
\textbf{Model} & \textbf{MAE} & \textbf{MAE\textsuperscript{M}} & \textbf{Accuracy} & \textbf{Weighted Accuracy} \\
\hline
ORNN & 0.769 (0.009) & 1.238 (0.016) & 0.818 (0.003) & 0.772 (0.004) \\
IT & 0.807 (0.010) & 1.244 (0.011) & 0.801 (0.003) & 0.781 (0.004) \\
AT & 0.778 (0.009) & 1.246 (0.012) & 0.813 (0.003) & 0.755 (0.004) \\
LAD & 0.829 (0.008) & 1.298 (0.016) & 0.786 (0.004) & 0.686 (0.003) \\
MC & 0.794 (0.012) & 1.286 (0.018) & 0.804 (0.003) & 0.767 (0.004) \\
HTDN & - & - & 0.800 & 0.753 \\
\hline
\end{tabular}
\end{table}

\noindent Here, IT: Isotonic Regression, AT: Additive Trees (likely gradient boosting), LAD: Least Absolute Deviation, MC: Multi-class Classification, and HTDN: Heteroscedastic Tobit Deep Network.

\textbf{Average Accuracies of LLMs Categorizing Posts from ReddiXNet}

\begin{itemize}
    \item GPT-4: 83.72\%
    \item Llama: 82.01\%
    \item Mistral: 83.29\%
    \item Gemini: 83.85\%
    \item Claude: 92.49\%
\end{itemize}

As we can see, the average accuracies by the best models of our paper, Diaz \& Panangadan (2020), and Wang et al. (2020) on their respective datasets are 92.49\%, $\sim$78\%, and 81.8\%, respectively. This shows that LLMs are performing better in the detection of such tasks.



\end{document}